\documentclass{report}
\usepackage{subfiles}

\begin{document}
\tableofcontents

\title{{Puff-like instability} in laminar to turbulence supercritical transition of round jets}%

\author{Neelakash Biswas$^1$,
  Aviral Sharma$^{2}$,
  Sandeep Saha$^{1*}$, 
 and Debopam Das$^{2}$} 

\email[]{Sandeep Saha: ssaha@aero.iitkgp.ac.in; \\
         Debopam Das: das@iitk.ac.in}

\affiliation{$^1$Department of Aerospace Engineering, Indian Institute of Technology Kharagpur, 721302, India}

\affiliation{$^2$Department of Aerospace Engineering, Indian Institute of Technology Kanpur, 208016, India}

\begin{abstract}
{We explore the laminar to turbulence transition of round jets at low Reynolds number ($Re < 1000$) using a novel experimental setup and linear stability theory (LST). \RevC{The setup has a large domain and a low disturbance environment which increases the critical $Re$ to approximately $500$,} \RevC{permitting the appearance of a hitherto unknown {\emph{puff-like instability} (PFI)}}. {The instability} is identified in the self-similar region of the jet through clean flow visualizations (FV) and further corroborated by particle image velocimetry (PIV) measurements. For $400<Re<700$, the flow exhibits \RevE{ \emph{PFI embedded in a puff-train} encapsulated by a laminar flow analogous to `puffs' in pipe-flow transition; the latter being symbolic of the \emph{finite-amplitude disturbance transition} scenario. The observation that PFI convects close to the average local velocity with an inflectional velocity profile further strengthens the analogy.} LST predicts that PFI is effectively a superposition of the helical mode (HM) pair with azimuthal wavenumbers $n = \pm 1$. Hence, {PFI} can also appear in the  \emph{infinitesimal-amplitude supercritical} route to transition of linearly unstable flows.}
\end{abstract}
{
\let\clearpage\relax
\maketitle
}
\hspace{7 pt} \textbf{\textit{Introduction}}.—  Jets are ubiquitous in nature, from the lengthscale of relativistic jets  \citep{meier2001magnetohydrodynamic}, to the jet stream \citep{joseph2003high} they prevail to the scale of human breathing \citep{tang2013airflow} and jellyfish propulsion \citep{dabiri2010wake}. Sophisticated engineering devices also require jets for diverse applications ranging from combustion \citep{nathan2006impacts} and propulsion \citep{levchenko2018recent} to drug delivery \citep{schramm2004needle}. Jets have received attention for a range of physical processes lately owing to their fascinating characteristics \citep{bejan2014evolution,cabanes2017laboratory,torres2020stars,underwood2019dynamic} and environmental impact \citep{mossa2016rethinking,mitsudera2018low}. Despite the interest of a broad community, the laminar to turbulence transition of jets remains relatively unexplored \citep{reynolds1962observations, viilu1962experimental,o2004stability}.  For instance, jets become turbulent in the near-field of the nozzle via either the helical mode (HM) or the Kelvin-Helmholtz instability \citep{yule1978large,verzicco1994direct,mattingly1974unstable}. {We report the \RevC{existence of}  \emph{puff-like instability} (PFI) \RevE{embedded in a \emph{puff-train} encapsulated by a laminar flow while  transitioning to turbulence \RevC{for the first time}, that bears startling visual semblance to turbulent puffs observed in transitional pipe flows \citep{mullin2011experimental,hof2010eliminating} (see SI section 1). Analogous to puffs in pipe flow, the PFI convects at nearly the local average velocity with an inflectional profile \citep{mullin2011experimental,hof2010eliminating}}}. \RevD{Indeed, the appearance of such transitional structures among canonical flows is ubiquitous: \emph{Puff-like structures (PFS)} in channel flow \citep{tsukahara2005dns} and \emph{turbulent spots} in boundary layers \citep{vinod2004pattern} too convect nearly at the same velocity. Whether their presence in wall- and un-bounded flows is a mere coincidence or indicative of a deeper connection between shear flows is a question of contemporary interest \citep{chantry2016turbulent,tuckerman2020patterns}.} \\
The appearance of \RevE{PFS} in transitional jets becomes even more profound in light of Landau's classification \citep{landau1944problem} of instabilities leading to transition. He suggested that transition could be triggered either by an \emph{infinitesimal-amplitude supercritical} perturbation leading to subsequent instabilities (for example, boundary layers) or by a \emph{finite-amplitude subcritical} perturbation (as observed in pipe or Couette flow). Puffs have been associated with the instabilities of the latter kind \citep{pomeau2016long}. Curiously in our case, the {PFI} belongs to the \emph{former} category. The {observed structures} are reminiscent of the arrow-headed turbulent spots in the supercritical transition of the boundary layer over a flat plate or an axisymmetric geometry \citep{vinod2004pattern,govindarajan_narasimha_2000}.

{The critical Reynolds number, $Re_c$, predicted by LST is $37.68$ for round jets \citep{batchelor1962analysis,lessen1973stability,morris1976spatial} ($Re \equiv {U_{av} D / \nu}$, $D, U_{av}, \nu$ denote nozzle diameter, average exit velocity and kinematic viscosity respectively). However, experimental investigations are yet to yield a concordant $Re_{c}$. For instance, \citet{viilu1962experimental} reported $Re_{c}$ $\sim$ $11$ whereas, Reynolds \cite{reynolds1962observations} reported the existence of long laminar jets for $Re<150$, whose length decreased gradually as Re approached 300; thereupon `confused breakdown'
ensued close to the exit ($\sim 100 D$)}. 

{Such discrepancy in $Re_c$ and jet breakdown-length reported in different experiments signify that jets remain laminar for long distances at high $Re$ depending upon the disturbance environment akin to pipe flows \cite{mullin2011experimental}. Indeed, we} observe laminar jets up to $1000 D$, even at $Re \sim 500$. Consequently, we are able to explore the far-field structures that were hitherto unknown and uncharacteristic of transitional round jets. \RevD{This result is profoundly relevant to investigations by \citet{tuckerman2020patterns} who report that numerical simulations in a large domain capture new structures in transition to turbulence similar to the large test section in our experiments.}

 Investigations using LST with \citep{morris1976spatial} and without \citep{batchelor1962analysis, mattingly1974unstable} the viscous effects showed that the eigenmodes are highly sensitive to the jet's base velocity profile. A `top-hat' profile near the nozzle exit {aids} the growth of both axisymmetric mode and HM {whereas,} the fully developed profile favours the latter {\emph{only}}. A parabolic profile, at the exit of a {long} pipe nozzle, is stable to axisymmetric disturbances \citep{kambe1969stability}. 
 The axisymmetric mode and the HMs have a region of absolute inviscid instability near the nozzle exit for lighter-than-ambient jets irrespective of the injector length \citep{coenen2008absolute}.
 Therefore, the theoretical results sum up to two important conclusions: a) the axisymmetric mode can be triggered in the near field either by a short nozzle or in a buoyant jet; b) the far field of the jet is stable to the axisymmetric mode. We perform `clean' flow visualizations (FV) and PIV experiments on low $Re$ jets emanating from a long pipe nozzle and report the structures present in the far field. In particular, we observe {PFI} far downstream ($x> 300D$). \RevC{Furthermore, LST predictions  quantitatively agree with the observed wavelength, arrowhead angle and the convection velocity of the PFI}.\\

\begin{figure}
\centerline{\includegraphics[clip = true, trim = 0 20 0 10,width= 0.43\textwidth]{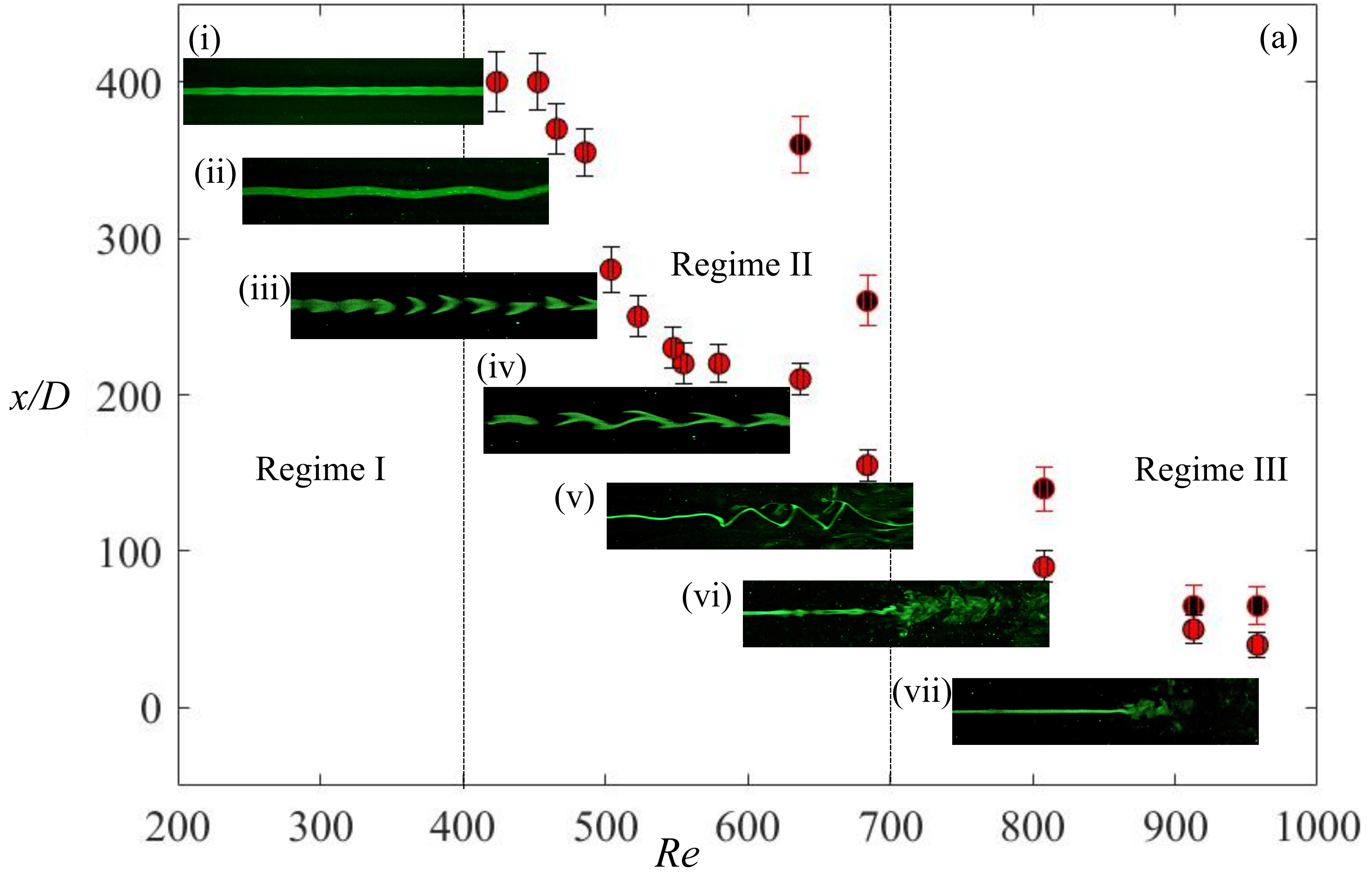}
\includegraphics[clip = true, trim = 0 -175 0 0,width= 0.0915\textwidth]{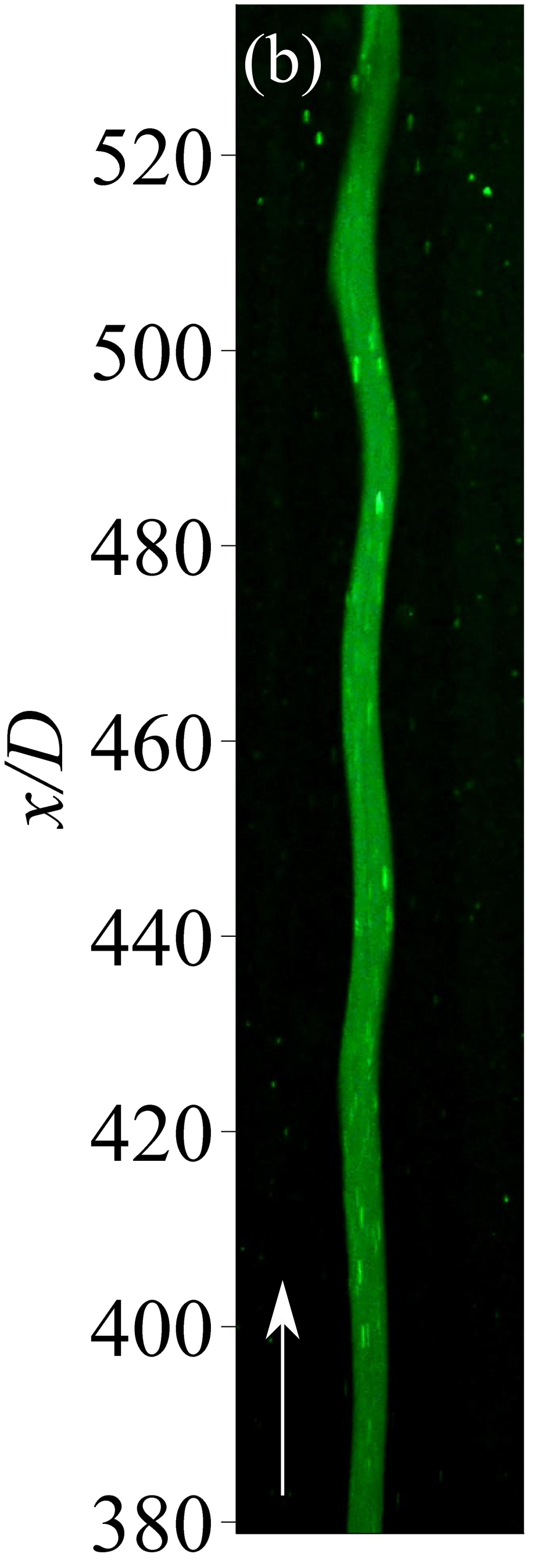}
\includegraphics[clip = true, trim = 0 -195 0 0,width= 0.080\textwidth]{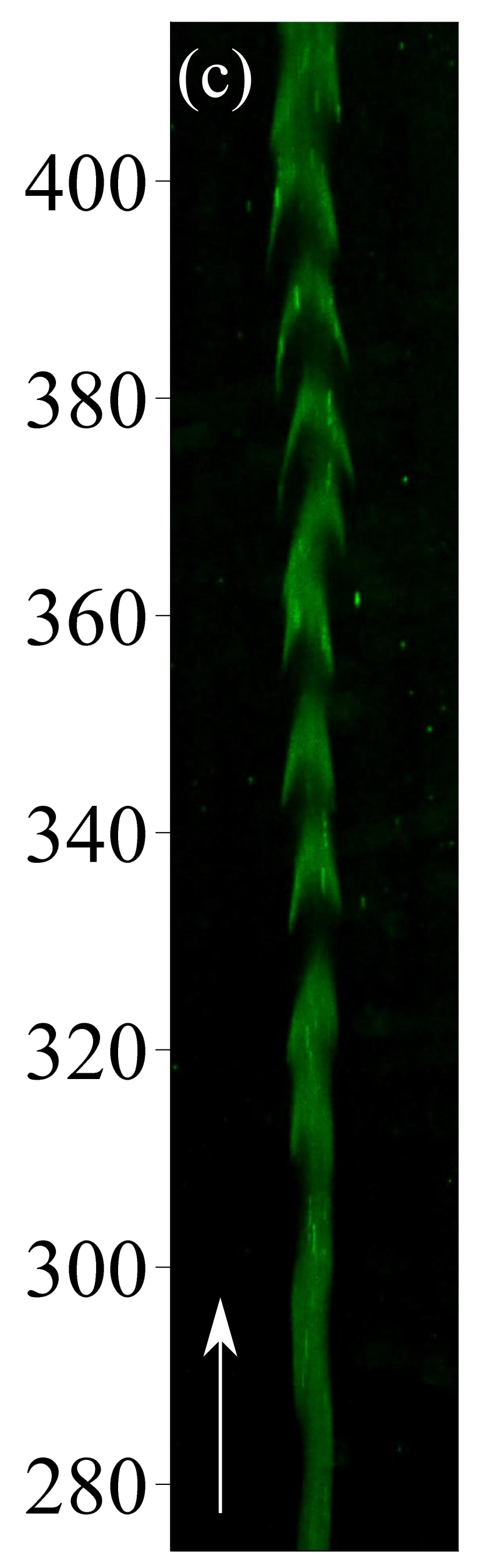}
\includegraphics[clip = true, trim = 110 90 50 0,width= 0.286\textwidth]{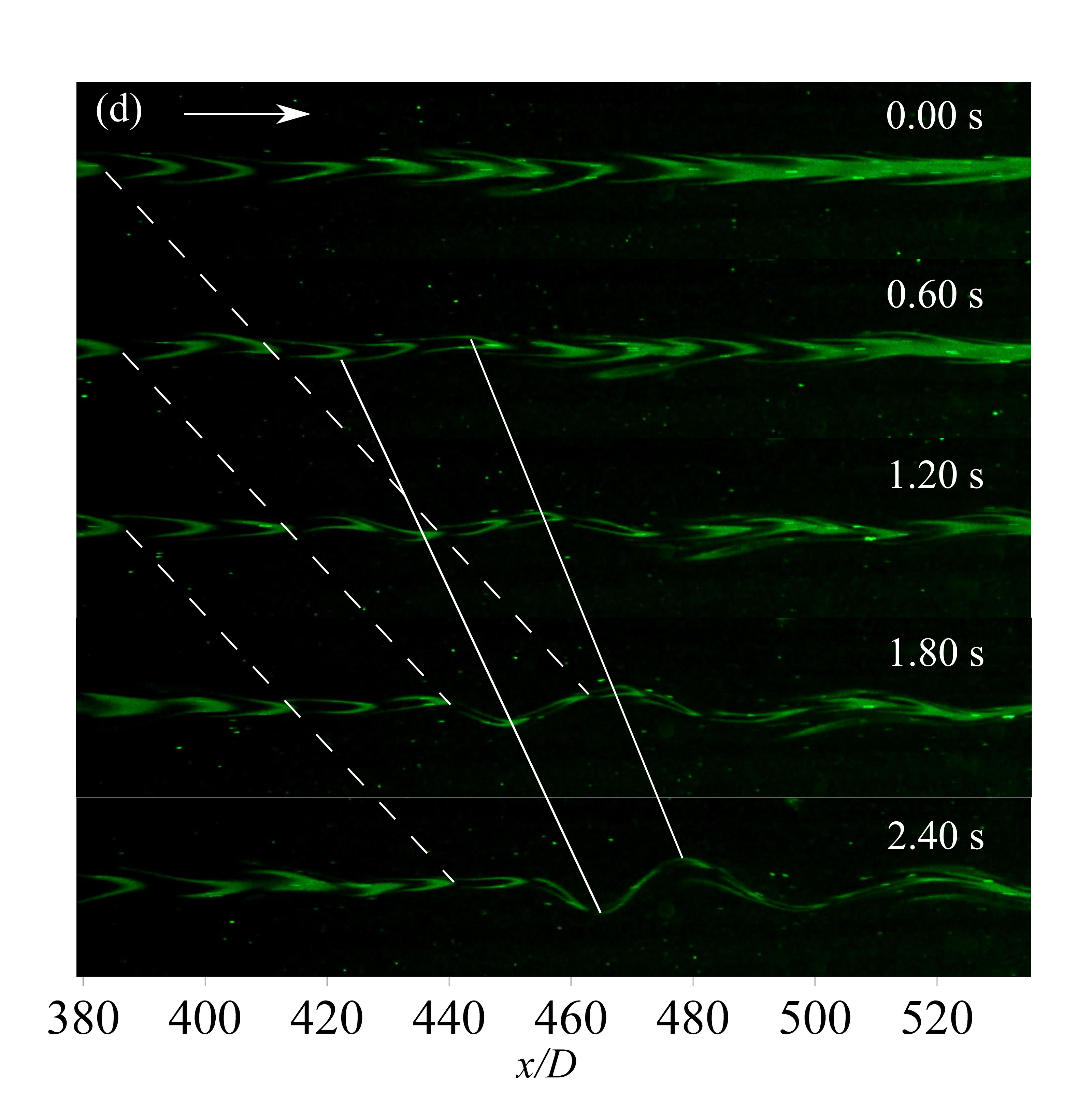}}
\caption{ (a) {Variation of minimum ( \RD{$\sbullet[1.2]$} $L_{min}$) {and maximum} ( \BLA{$\sbullet[1.2]$} $L_{max})$ jet breakdown length from the nozzle exit with $Re$}. The variation of the structure of the jet instability with increasing Re is shown in insets at  $Re =$ (i) $395$ ;(ii) $434$; (iii) $459$; (iv) $489$; (v) $600$; (vi) $700$; (vii) $850$. (b) HM at  $Re = 434$ and the (c) {PFI} as seen at $Re = 435$. The jet discharges horizontally and the white arrows show the direction of flow. (d) Spatio-temporal evolution of the HM and {PFI} at $Re  = 443$.  {Location of consecutive {fronts of PFI}  (\protect\tikz[baseline]{\protect\draw[line width=0.2mm,loosely dashed] (0,0.8 mm)--+(1.08,0)}) and  the crest and trough of the helical wave (\protect\tikz[baseline]{\protect\draw[line width=0.2mm] (0,0.8 mm)--+(1.08,0)}) are shown whose slope indicates the convection velocity.} See SI for details of the experimental setup and supplementary movie 1 shows the instabilities.}
\label{fig:kd}
\end{figure}

 
\hspace*{7 pt} \textbf{\textit{Experimental setup}}.— \RevC{We fabricated a novel, low-disturbance-environment water jet setup (see SI section 2). 
A horizontal jet discharged into a large tank with an overflow chamber to prevent mass accumulation, thus eliminating back-pressure fluctuations and recirculation. A Mariotte bottle arrangement ensured a constant head discharge. Inlet disturbances were further suppressed using a settling chamber comprising of a sponge and a bell-mouthed nozzle entry. A nozzle length to diameter ratio ($L/D$) of $300 - 450$ ensured the parabolic profile at the highest $Re$ considered.}\\

\begin{figure}
  \centerline{  \includegraphics[clip = true, trim = 450 40 600 70,width= 0.12\textwidth]{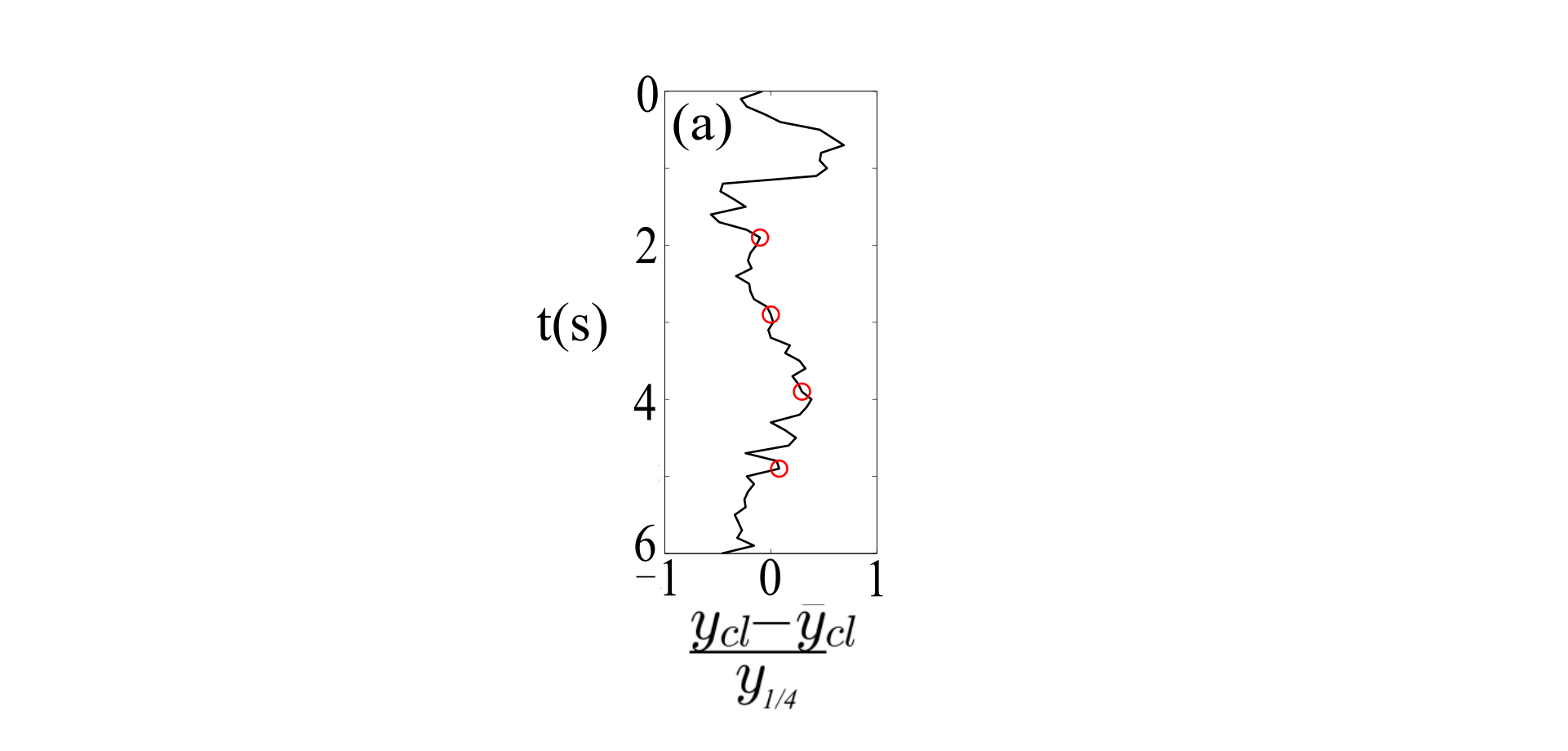}
  \includegraphics[clip = true, trim = 0 5 0 9,width= 0.315\textwidth]{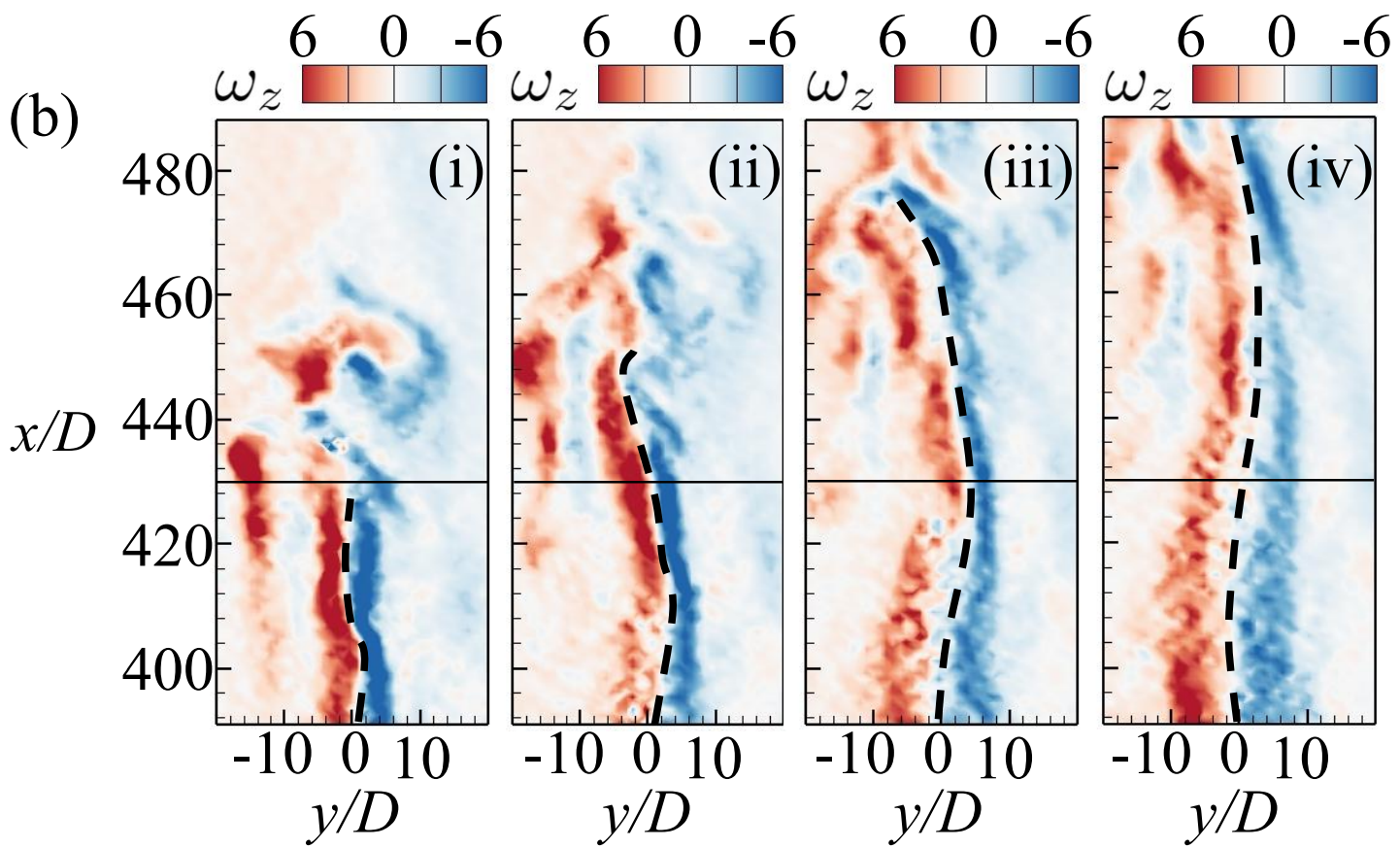}
  \includegraphics[clip = true, trim = -25 10 15 17,width= 0.23\textwidth]{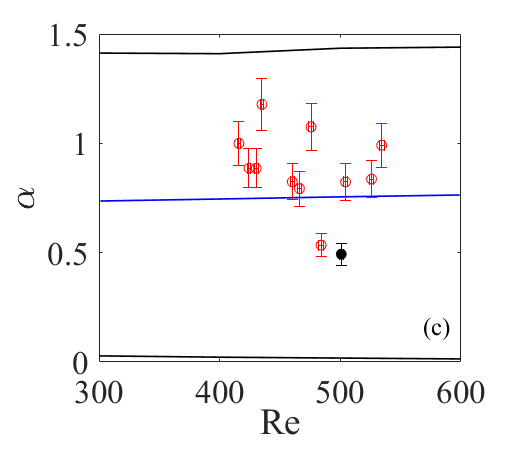}
  \includegraphics[clip = true, trim = 0 10 15 20,width= 0.215\textwidth]{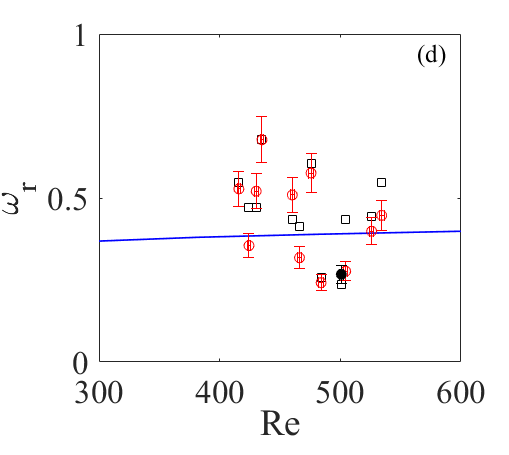}}
 \caption{(a) Nondimensional jet-centerline fluctuation over time for $Re = 500$ at $x/D = 430$, shown by the horizontal line in Fig.\ref{fig:kt}(b). (b) Vorticity ($1/s$) contours at the points (\RD{$\circ$}) of Fig. \ref{fig:kt}(a). The dashed line tracks the jet-centerline. (c) HM wavenumbers, FV (\RD{$\circ$}), PIV ($\sbullet[1.2]$), dominant HM from LST (\NB{\solid}). Neutral curve of HM from LST (\BLA{\solid}). (d) HM frequency ($\omega_r$) from FV (\RD{$\circ$}), PIV ($\sbullet[1.2]$) and dominant mode from LST  (\NB{\solid}). The frequencies obtained from LST corresponding to the experimentally obtained wavenumbers (Fig. \ref{fig:kt}(c)) are shown by ($\square$).}
\label{fig:kt}
\end{figure}

\hspace*{7 pt} \textbf{\textit{Flow Visualization}}.— Dye visualizations (see SI section 3) were carried out from $Re \sim 200 - 1000$. \RevD{The estimated uncertainty in $Re$ was within $0.25\%$  of $Re$ (see SI section 8)}. {The location of inception and the initial amplitude of the instability are prescribed by the precise disturbance environment and impacts the jet's length at breakdown. The maximum ($L_{max}$) and the minimum ($L_{min}$) length of the jet, at  breakdown in different realizations of the same experiment are plotted in Fig. \ref{fig:kd}(a) and the associated structures in the insets.} Three flow regimes are delineated: In regime I ($Re < 400$), no breakdown is observed up until $1000 D$. Thereupon, Regime II ($400< Re <700$) ensues exhibiting a range of flow structures hitherto unreported (Fig. \ref{fig:kd}(a) inset (ii) - (v)). \RevD{Very long ($L_{max }\approx 1000D$) laminar jets  are still encountered upto $Re \approx 500$ in Regime II in some experiments depending on the background disturbance levels. Thus, $Re_c \approx 500$ in our study is significantly higher than  previous experimental studies \citep{viilu1962experimental, reynolds1962observations, o2004stability}, a \emph{direct consequence} of the low background disturbance level in the experimental setup. In Regime II, the HM (shown at $Re \approx 434$ in Fig. \ref{fig:kd}(a) (ii)) often alternates with the PFI (shown at $Re \approx 459$ (Fig. \ref{fig:kd}(a)(iii)) and the sequence of events varied between experiments. The coalescence of the two structures is shown at $Re \approx 489$ in Fig. \ref{fig:kd}(a)(iv). Gradually the HM supersedes the {PFI} as we reach higher $Re \approx 600$ (inset Fig. \ref{fig:kd}(a)(v)) eventually leading to breakdown.} Finally, in Regime III, ($Re > 700$), the jet transitions to turbulence abruptly affirmed by $L_{max}$ and $L_{min}$ being minimal and nearly equal. The focus of the article is Regime II where uncharacteristic flow structures appear unlike Regimes I and III. The HM and {the PFS} observed in Regime II are shown in Fig. \ref{fig:kd} (b-c). \BB{The HM is dominant in the fully developed region and prevails throughout Regime II. \RevD{Contrarily, the {PFS} were relatively less frequent in the fully developed region and were observable only in a limited range $400<Re<550$ alike puffs in pipe flows ($Re \in 1760-3000$)} \cite{mullin2011experimental}. All the {PFS} had an arrowhead pointing downstream with a long tail at a half-angle between $ 8^\circ$ to $15^\circ$ with the centerline.} The spatio-temporal evolution of the {PFS} at $Re = 443$ is shown in Fig. \ref{fig:kd}(d). The PFS coalesce within themselves as they convect downstream and occasionally co-exist with the helical undulations {(Fig. \ref{fig:kd}(d), $t=1.8s$).} We track the fronts of the PFS and the crest and trough of the helical wave to estimate the convective velocity. The slope of the lines reveals that the helical wave travels much slower than the PFS. The convection velocity of the PFS and the HM is approximately  $4.4 cm/s$ and $2.4 cm/s$ respectively.\\ 

\begin{figure}
  \centerline{ \includegraphics[clip = true, trim = 0 10 0 0,width= 0.32\textwidth]{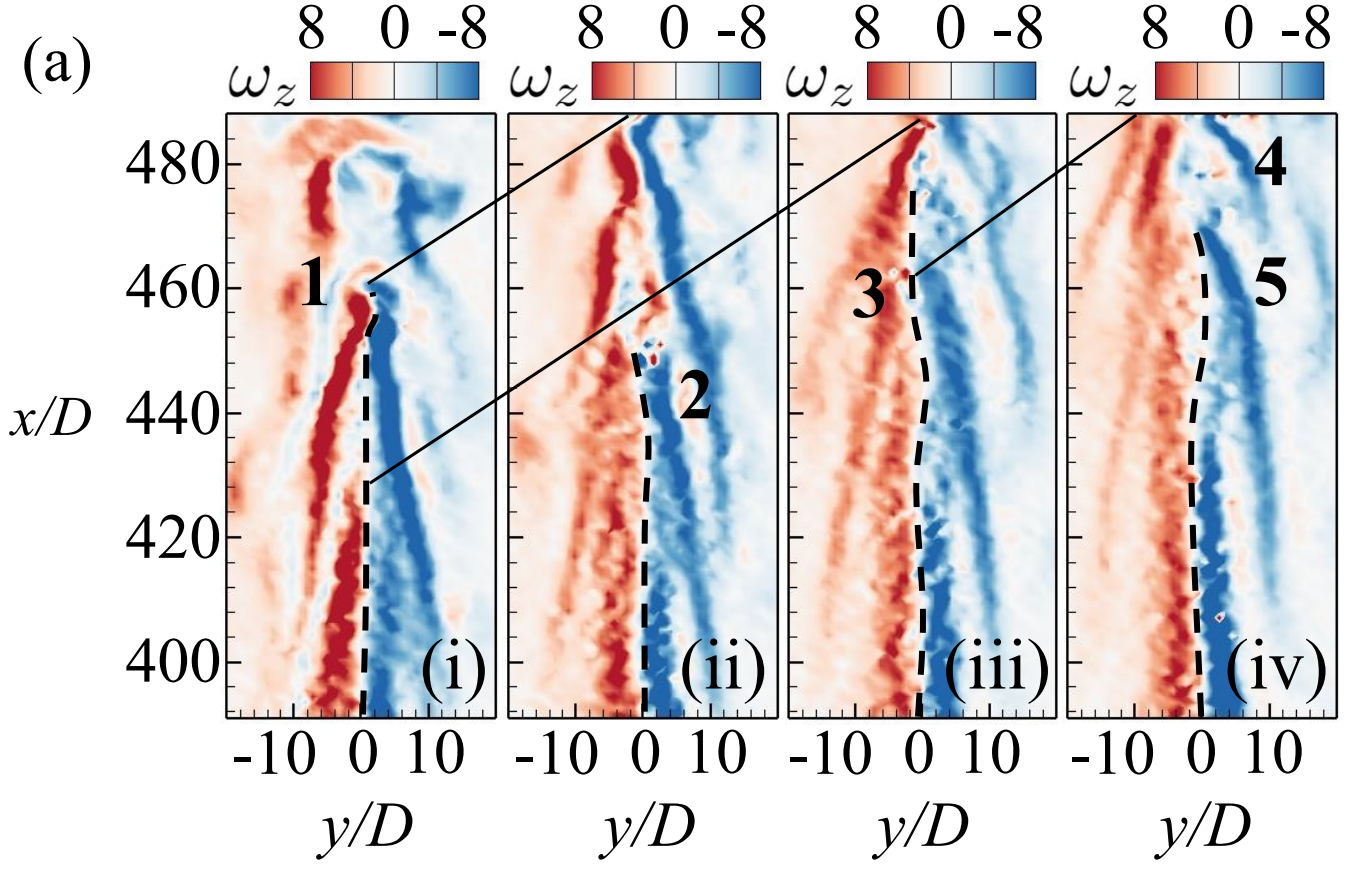}
  \includegraphics[clip = true, trim = 1700 350 1500 0,width= 0.105\textwidth]{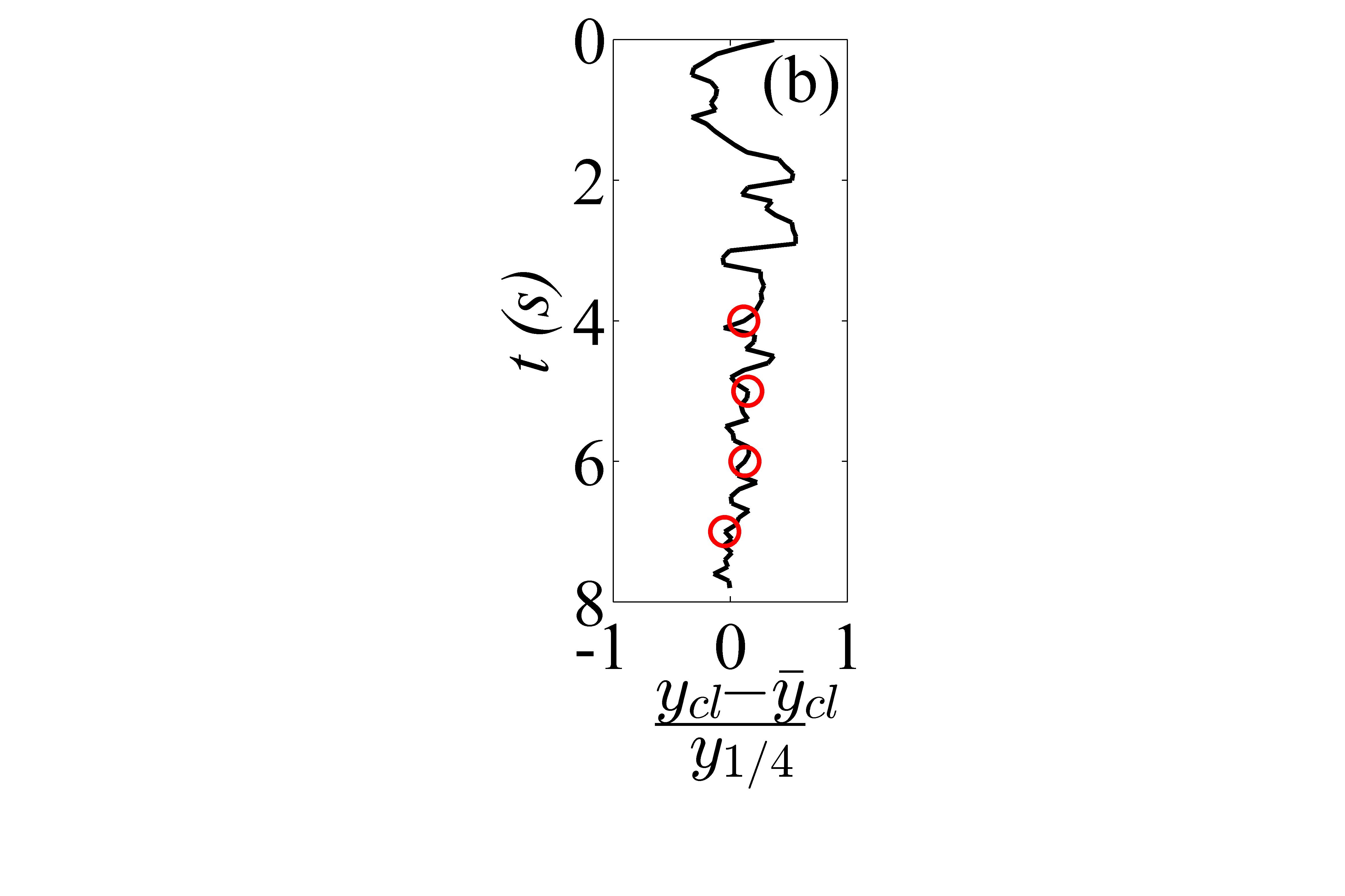}}
 \caption{ (a) Evolution of PFS at $Re = 550$. Vorticity ($1/s$) contours are shown at an interval of $1 s$. The PFS are numbered chronologically. (b) Jet-centerline fluctuation in presence of PFS at $x/D = 440$. }
\label{fig:ktt}
\end{figure}

\begin{figure*}[!ht]
  \centerline{\includegraphics[clip = true, trim = 0 0 0 0 ,width= 0.3\textwidth]{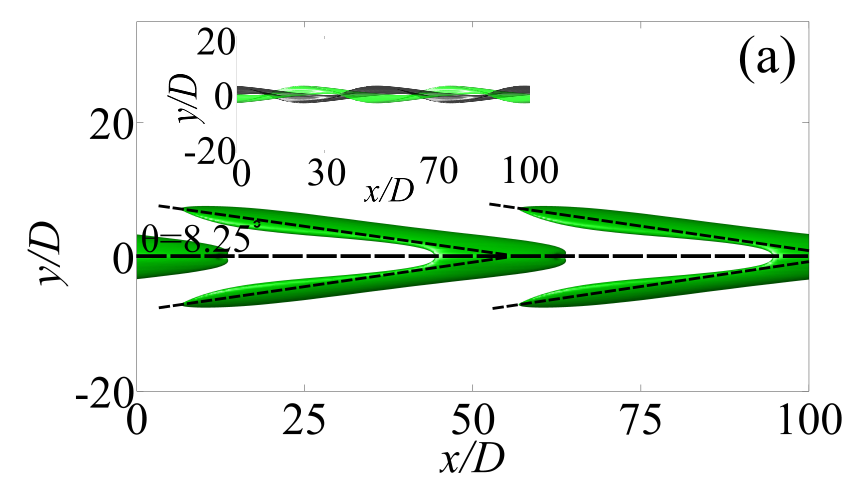}
  \includegraphics[clip = true, trim = -10 0 -50 0 ,width= 0.32\textwidth]{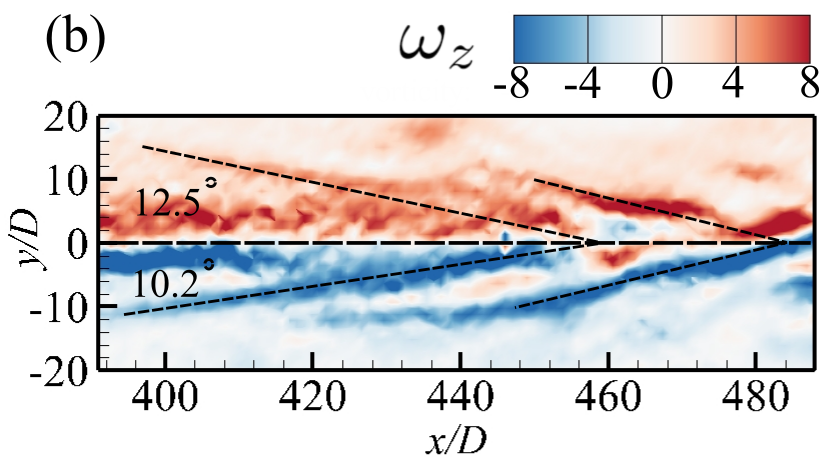}
  \includegraphics[clip = true, trim =0 0 0 0 ,width= 0.29\textwidth]{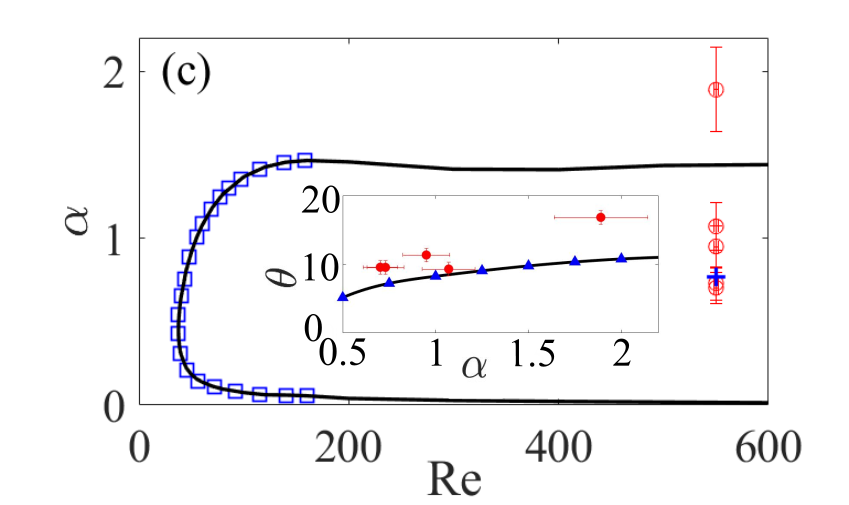}}
 \caption{ \BB{(a) Isosurface of $\omega_z = 2$  obtained by superimposing $n=\pm 1$ modes at $\alpha = 1$. Inset: Isosurface of $\omega_x = 2$ for $n=1$ (green) and $n=-1$ (black) modes showing their helical nature. (b) A PIV snapshot showing PFS at $\alpha \sim 1$. (c) The open circles (\RD{$\circ$}) show the wavenumbers of the PFS obtained through PIV. The sign (\NB{+}) shows the most unstable wavenumber of $n=\pm 1$ mode. Inset: The variation of experimentally observed $\theta$ with $\alpha$ (\RD{$\sbullet[1.2]$}) compared to LST prediction (-\tikztriangleleft[blue!60,fill=blue!50]-).}}
\label{fig:kl}
\end{figure*}

\hspace{7 pt} \textbf{\textit{PIV and LST}}.— Coexisting HM and PFS in the far field of the jet are also observed in PIV measurements. The planar PIV experiments were conducted at $Re = 450, 500, 550, 650$ with the field of view between $x/D$, $390-490$. Concordant with the FV results, PIV data also show that the jet remains laminar throughout the field of view (upto $x/D \approx 490$) even at $Re \approx 500$. The measured velocity profile corresponding to a fully-developed laminar jet agrees quantitatively with the laminar solution (see SI section 4). 
\\
We conduct a temporal linear stability analysis \cite{lessen1973stability} using the laminar solution \citep[pp 81-83]{landau1959fluid} as the base velocity profile (see SI section 4). The HM manifests itself in terms of an undulation of the jet centerline in the central diametrical plane; a signature that is absent when PFS appear solely.  A time signal of the jet-centerline fluctuation  ($\frac{y_{cl}-\bar{y}_{cl}}{y_{1/4}}$) at $Re = 500$ and $x/D = 430$ shown in Fig. \ref{fig:kt}(a) reveals how the onset of the helical instability affects the jet-centreline location. The time averaged jet-centerline location ($\bar{y}_{cl}$) has been subtracted from the instantaneous centerline ($y_{cl}$) location and nondimensionalized using local quarter-velocity width ($y_{1/4}$) to compute the fluctuations. In order to ascertain that the signal indeed corresponds to the HM, we compare the  nondimensional spatial wavenumber ($\alpha$)  and the temporal  frequency $\omega_r$ obtained experimentally with LST. The reference length and velocity scale are $y_{1/4}$ and the centreline velocity, $U_{cl}$ respectively.  A Fast Fourier transform (FFT) of the signal reveals that the dominant frequency is around $0.3Hz$ {(see SI section 7)}. Vorticity contours of the flow field are shown in Fig. \ref{fig:kt}(b) at instances corresponding to the equally spaced circles on the signal. A spatially undulating nature of the jet centerline is evident. The average wavelength of the undulation is approximately $10${$cm$}, which corresponds to $\alpha \sim 0.5$. Also, the dominant frequency corresponds to $\omega_r = 0.27$. At this $Re$ and $\alpha$, LST predicts the HM is unstable with $\omega_r\sim 0.2352$, which is close to the experimental observation. {The set of wavenumbers and frequencies of the HM obtained from FV and PIV experiments are shown in Fig. \ref{fig:kt}(c) and Fig. \ref{fig:kt}(d) respectively. The experimentally obtained $\alpha$, $\omega_r$ of the HM are clustered near the dominant mode predicted by LST (\NB{\solid}) and are well within the unstable region of the neutral curve (\BLA{\solid}).} Spatio-temporal evolution of the flow-structures at $Re = 550$ is shown in Fig. \ref{fig:ktt}(a). The arrowhead shaped structures with long tails resemble the PFS shown in Fig. \ref{fig:kd}(c) and (d) and the associated jet-centerline fluctuations are lower than the HM in Fig. \ref{fig:kt}(a-b). Although all the PFS convect nearly at the same velocity ($\sim 3cm/s$), their wavelength varies between $2.5cm$ ($\alpha \sim 1.8$ for PFS 4) to $6.5cm$ ($\alpha$ $\sim$ 0.7 for the structures 3 and 5). The half angle, $\theta$ (see SI section 6), between the tail of the PFS and the centerline, is always small, $9.3^\circ, 11.4^\circ, 9.6^\circ, 16.8^\circ, 9.6^\circ$ for {structures} 1-5 respectively. \RevD{Coincidentally, the half-angle subtended by the arrow-headed turbulent spots at its origin in zero pressure gradient boundary layers is approximately $10^\circ$ \citep{vinod2004pattern}, and varies by a few degrees depending upon the applied pressure gradient \citep{narasimha1985laminar}}. Similar to Fig. \ref{fig:kd}(d), a co-existence of helical undulation along with the PFS is observed in \ref{fig:ktt}(a) at initial instants of time. The jet-centerline location in presence of the PFS shows much weaker undulations in Fig. \ref{fig:ktt}(a) in comparison to Fig. \ref{fig:kt}(b). The jet-centerline fluctuation is shown in Fig. \ref{fig:ktt}(b) and the circles on the plot represent the instants at which vorticity contours are shown in Fig. \ref{fig:ktt}(a). The signal shows the presence of strong fluctuations in the jet prior to the arrival of the PFS. The fluctuations subside markedly with the passage of the PFS as shown by the diminishing undulations of the centerline in Fig. \ref{fig:ktt}(b) (and Fig. 8 of SI). \\

\hspace{7 pt} \textbf{\textit{On the origin of PFS}}.— 
\RevC{LST predicts that $n = \pm 1$ are the only two unstable modes and hence a  \emph{non-linear} interaction of these two modes gives rise to a possibility of creating a robust far-field structure like PFS (SI Fig. 8). A non-linear model could illustrate complex interactions like phase synchronization and saturation \cite{pomeau2015transition,pomeau1986front,leweke1995flow}. However, significant insight may be obtained based on a linear superposition \citep{yoda1994instantaneous} of the unstable modes, which we justify \textit{a posteriori}.} We superpose the HM pair $n=\pm1$ for $\alpha = 1$ at $Re = 550$. {The streamwise component ($\omega_x$) of the HMs \citep{danaila1997coherent,bose2016helical} demonstrates the spiraling nature visually in the inset of the Fig. \ref{fig:kl}(a). The iso-surface of spanwise vorticity, $\omega_z$ (obtained from LST) has a conical shape in Fig. \ref{fig:kl}(a) and resembles the spanwise vorticity field of the PFS (from PIV) in Fig. \ref{fig:kl}(b) at the same $Re$ having similar wavelength.} The half angle of the PFS, $12.5^\circ$ and $10.2^\circ$ are also reasonably close to the half angle, $8.3^\circ$, predicted by LST. Furthermore, $\alpha$ of all the observed PFS (\RD{$\circ$}), except for one, are close to the most unstable wavenumber (\NB{+}) in Fig. \ref{fig:kl}(c) and lie within the unstable region of the HM. The half angle predicted by LST  {(see SI section 6)} and PIV also match closely (inset of Fig. \ref{fig:kl}(c)). \RevC{The difference between predicted (LST) and observed $\theta$ is likely to be a consequence of the non-linear effects. The equal growth rate (Fig. 9(b) of SI) of the HMs ($n = \pm 1$) plausibly allows a non-linear interaction of the mode pair to form the PFS.}



\begin{figure}[!ht]
  \centerline{
  \includegraphics[clip = true, trim = 0 13 0 0 ,width= 0.375\textwidth]{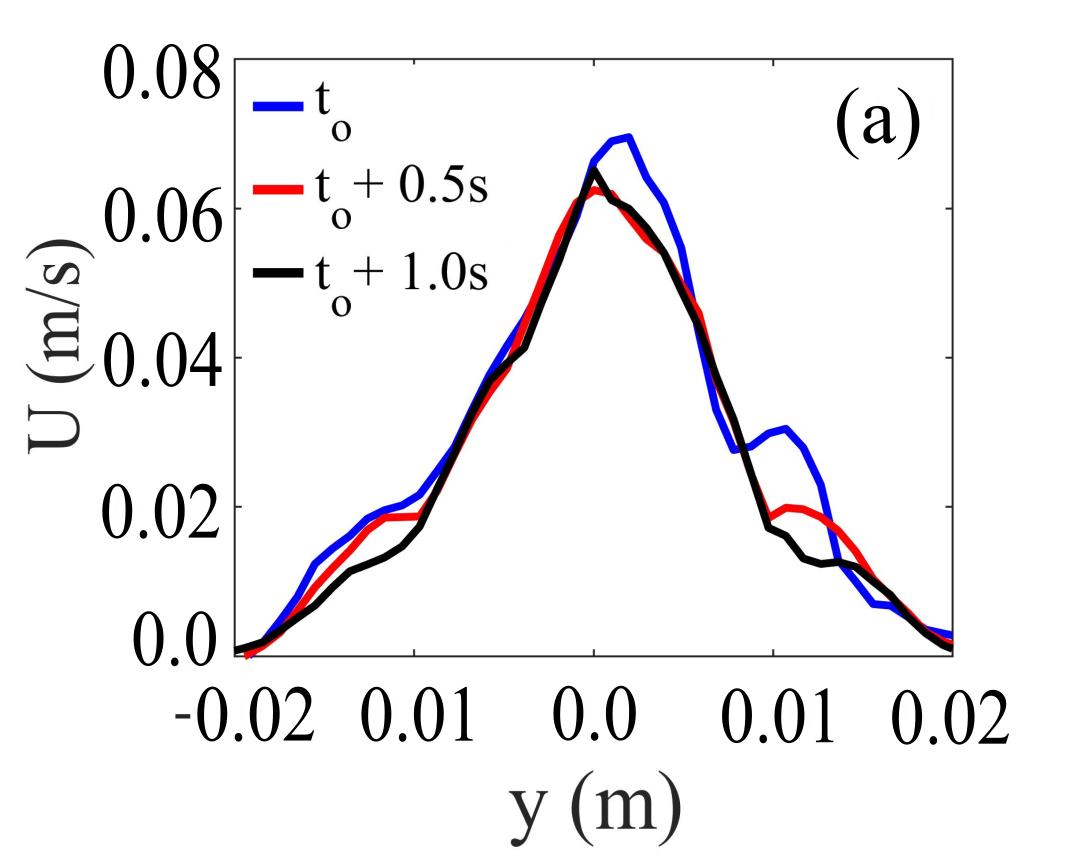}
  \includegraphics[clip = true, trim = 0 10 0 15 ,width= 0.3525\textwidth]{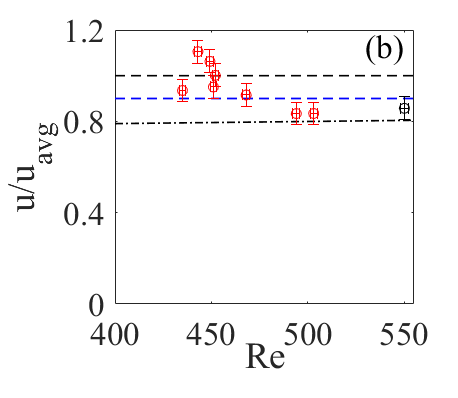}
    }
 \caption{(a) Instantaneous velocity profiles in presence of PFI at $x/D=435$. Time $t=t_o$ corresponds to Fig. \ref{fig:ktt}(a)(ii) (b) Convection velocity of PFI obtained from FV (\RD{$\circ$}) and PIV ($\circ$). Convection velocity of puffs in pipe flow (\BLA{\textbf{- -}} \citep{hof2010eliminating}, \NB{\textbf{- -}} \citep{mullin2011experimental}). Convection velocity of dominant HM from LST (\textbf{-- $\cdot$ --}).}
\label{fig:puff_velocity}
\end{figure}

{The puffs in pipe flow have been attributed to the inflection points in the local velocity profile (see inset of Fig. 1A of \citet{hof2010eliminating}) and similar inflection points can be observed in the instantaneous velocity profiles in Fig. \ref{fig:puff_velocity}(a). Generation of streamwise vorticity has been reported in pipe flow close to the inflection points \citep{hof2010eliminating}. Similarly, the HMs are a source of streamwise vorticity as observed in the inset of Fig. \ref{fig:kl}(a). The connection  is further strengthened by the observation that the PFI convects approximately (see SI section 5) at the average local velocity (see Fig. \ref{fig:puff_velocity}(b)), similar to the puffs in pipe flows \citep{mullin2011experimental,hof2010eliminating}. In addition, the LST prediction also matches this result closely (Fig. \ref{fig:puff_velocity}(b)).  The link prompts future investigation of how ``puff-like'' arrow-headed structures appear in transition of both, wall-bounded and free-shear flows \citep{tuckerman2020patterns, chantry2016turbulent}.
} \\

\hspace{7 pt} \textbf{\textit{Conclusions}}.— \RevC{We carried out clean flow visualization and PIV experiments in a large domain having a low disturbance environment and obtained} long and stable laminar jets at Re ($\approx 500$) significantly higher than the $Re_c$ reported previously \citep{viilu1962experimental, reynolds1962observations, o2004stability}.
We discovered the existence of \RevE{a train of \emph{Puff-like instabilities} bounded by a laminar flow}, which is represented as a superposition of the HM pair ($n = \pm1$). The lengthscale and the half angle of the PFI are largely described by a \emph{supercritical infinitesimal perturbation} \RevC{although the structures themselves are associated with non-linear fluctuations.}  
The presence of PFI in supercritical transitional jets having similarities with `puffs' in pipe flows is thus startling, because the latter are representative of the \emph{finite-amplitude disturbance} route to subcritical transition in the sense of Landau \citep{landau1944problem}. \RevD{Further, the similarities between \RevE{puff-train} and `puffs' in pipe flows, which connect transitional free-shear and wall-bounded flows in a large domain merit future investigation in light of recent results of \citet{tuckerman2020patterns}.}  \\

\textbf{{Acknowledgements}:} We are grateful to the reviewers for key inputs that enriched the manuscript substantially.\\

 \textbf{\textit{Supplementary information}}.— \\

\noindent \textit{{1. Puff-train embedded in a laminar background}}\\
  
\begin{figure}[h]
    \centering
    \includegraphics[width = 0.55\textwidth]{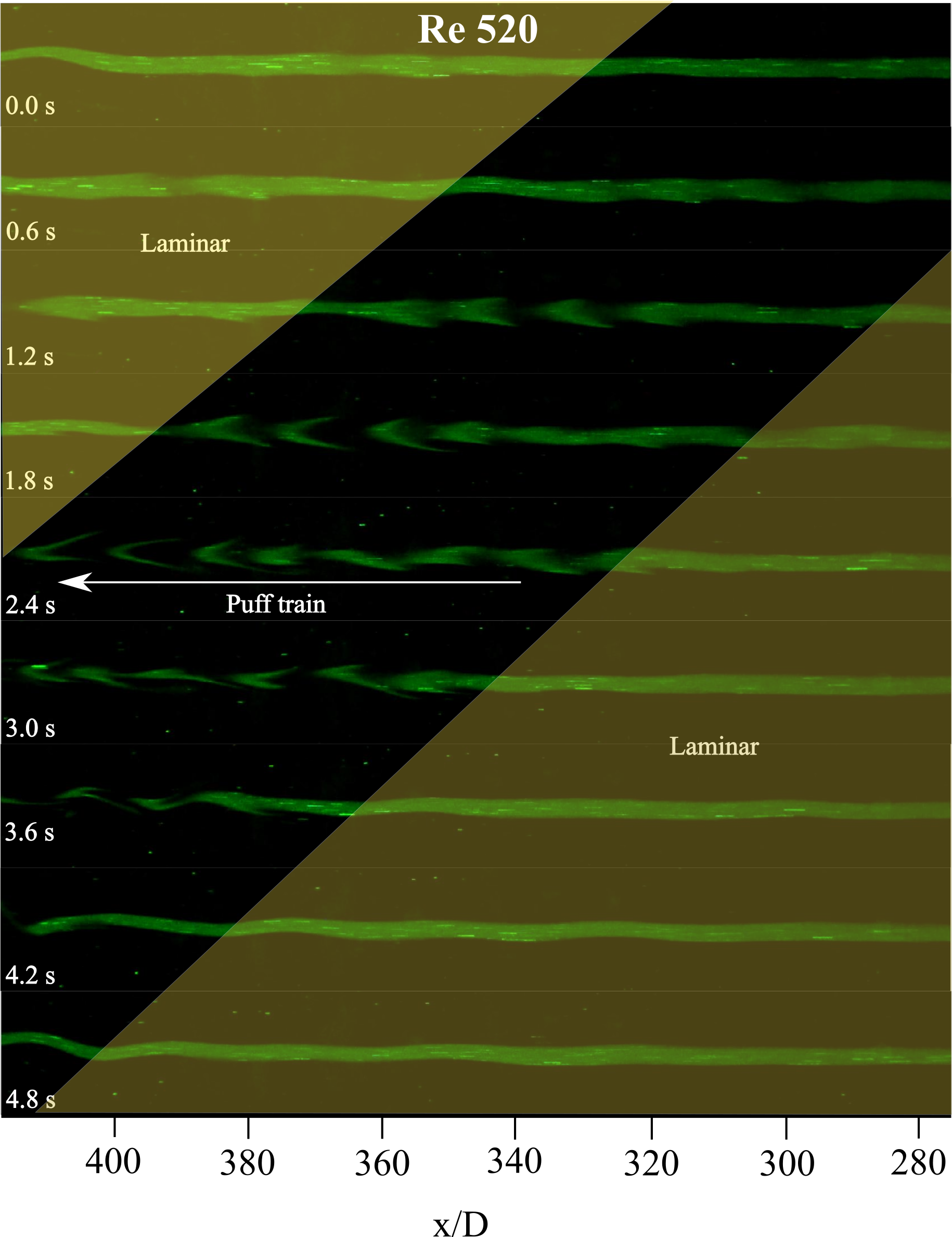}
    \caption{Arrow-headed Puff like structures constituting a solitary puff-train embedded in laminar filament at $Re = 520$.}
    \label{fig:puff_train}
\end{figure}

    \RevE{Figure \ref{fig:puff_train} demonstrates a solitary puff-train (akin to puffs in pipe flow) embedded in a laminar background at $Re = 520$ (similar puff-trains are observed at other Re). Here, $Re \equiv {U_{av} D / \nu}$, $D, U_{av}, \nu$ denote nozzle diameter, average exit velocity and kinematic viscosity respectively. Similar solitary wave-trains embedded in laminar flows resembling puffs in pipe flows exist in two dimensional Poiseuille flows \citep{jimenez1990transition}. The existence of a solitary puff-train bolsters the connection between \emph{puff-like structures} and puffs in transitional pipe flows.} \\
    
\noindent \textit{{{2. Experimental setup}}}\\
 
\begin{figure}[!ht]\vspace*{-0.15truein}
  \centerline{\includegraphics[clip = true, trim = 0 0 0 -20 ,width= 0.9\textwidth]{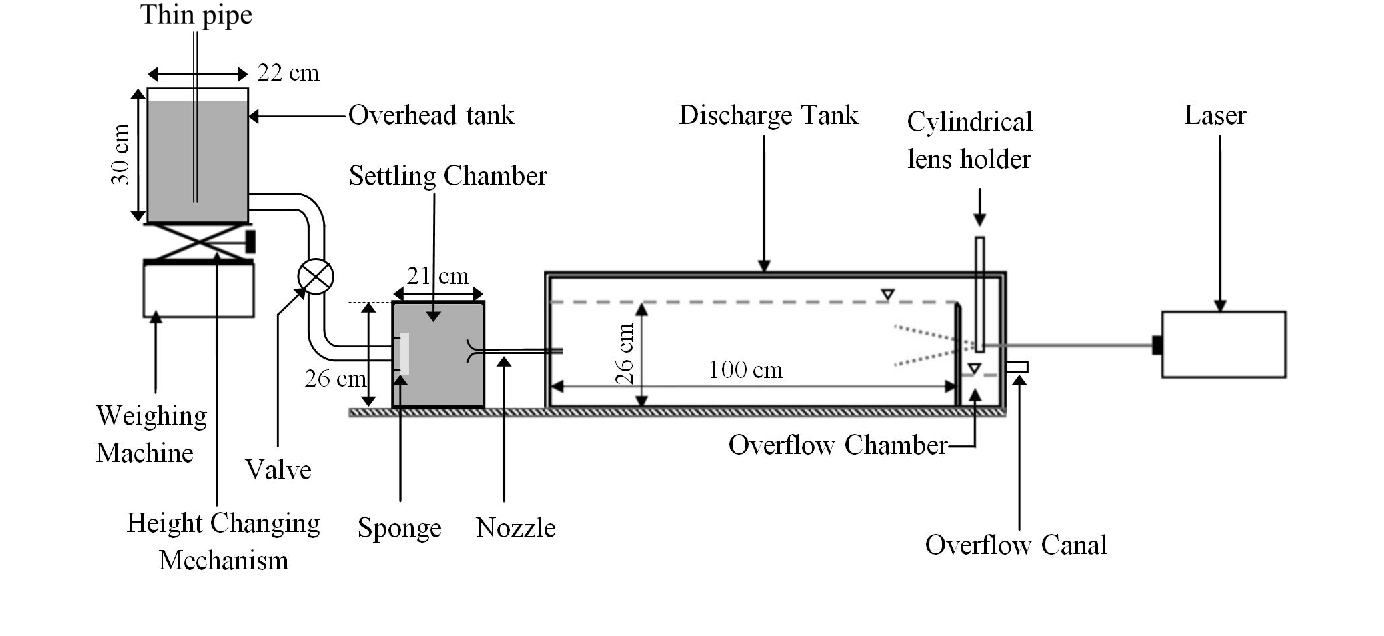}}
   \caption{ Schematic of the experimental setup. }
\label{fig:SIsetup}
\end{figure}

 \noindent A detailed schematic of the experimental setup is shown in Fig. \ref{fig:SIsetup}. A pipe nozzle of length $L$ and inner diameter ($D$) 1$mm$ is used in a \emph{large} discharge tank (dimension $1.2\times0.3\times0.3$ $m^3$)  to generate the jet. The $L/D$ ratios of $300$ and $ 450$ are used for flow visualization and the PIV experiments respectively to ensure that the exit velocity profile is fully developed. The dimensions of the overhead tank, settling chamber and discharge tank is shown in the schematic. The jet is drawn from the overhead tank, into which a thin pipe is inserted to use the principle of Mariotte's bottle and obtain a constant head over the duration of the experiment. Thus, a steady flow rate for time periods much longer than the observation period (typically around 1-2 minutes) is achieved. The overhead tank height is set using a height-changing mechanism as shown in the schematic in order to control the volume flow rate. The volume flow rate is measured accurately by placing the overhead tank along with the height-changing mechanism on a precision weighing machine with least count $0.01~\hbox{gram}$. Measuring the change in the weight of the overhead tank over a specified period of time provided the volume flow rate. 
 
 
 At the beginning of each experiment the water level is kept at the level of over-flow chamber barrier and thus, as the jet flow starts, the empty overflow chamber starts to fill avoiding any mass accumulation in the discharge tank. The temperature of the water inside the discharge tank is noted up to two decimal points using a thermocouple probe during an experiment. The Mariotte's bottle being sealed from the top and the discharge tank being covered prevent any evaporation. The experimental facility is kept in a large quiescent laboratory without any air-conditioning to avoid any disturbances due to flow from the air-conditioner. Further, the experiments are conducted between late night and early morning to minimize the ambient external disturbances and human movement near the experimental facility is restricted during an experiment.

\RevC{To achieve low disturbance environment we incorporate a range of novel measures in the experimental setup. A settling chamber is placed between the overhead and the discharge tank. A sponge-layer at the inlet of the settling chamber suppressed small disturbances entering the settling chamber. A bell-mouthed inlet of the pipe nozzle minimised the inlet disturbances. The velocity at the entry of the settling chamber is less than $50mm/s$ and a smooth entry of the flow is verified using a dye stream visualization. \citet{hussein1994velocity} had demonstrated that the jet at $Re \approx 10^5$ closely resembles a jet in an infinite environment using an enclosure $984 D$ in length. Hence the effect of flow recirculation was expected to be minimal at a much lower $Re \sim 10^3$ with a discharge tank length of $1000D$ in present experiments. The overflow chamber maintained a constant hydrostatic head, limited the influence of back-pressure and consequently eliminated almost any recirculation.} \\
 
 \noindent \textit{{3. Experimental methodology}}\\

 \noindent Experiments are carried out with filtered water. During flow visualization, $20 ml$ of Sodium Fluorescein dye solution is added to 5 litre water. For flow visualization experiments, the flow field is{ illuminated} with a continuous laser (1.2W, 532nm). Images are captured using Nikon D5200 camera and videos are recorded with Sony HDR-SR10E at 25fps. For PIV experiments, a double-pulsed, Nd:YAG laser having energy of $200mJ/pulse$ and repetition rate of $10Hz$ is used. The light is delivered from the laser head to the flow field under investigation through an articulated arm with a sheet forming optics, which consists of two spherical lenses and a cylindrical lens. The sheet is approximately $1mm$ thick and lies vertically in the central diametral plane of the jet. A $12$-bit, charge coupled device (CCD) camera of resolution $3312\times2488$ pixels is used at $10Hz$ frame rate in double frame mode. Hollow glass spheres of average diameter $10 \mu m$ are used as seeding particles which are uniformly mixed with the water in the discharge tank and kept for nearly an hour before every experiment to ensure that the eddies in the discharge tank die down and do not affect the jet. The Stokes number of the particles in this flow is of the order of $10^{-5}$  \citep{das2017generation} and thus expected to trace the flow faithfully. Image pairs are obtained at 10$Hz$ frequency using MICROVEC PIV system and average particle displacement was calculated using PIV lab \cite{thielicke2014pivlab}. The adaptive correlation algorithm in PIVlab used a multi-pass, fast Fourier transform to determine the average particle displacement.  An initial interrogation area (IA) of $256\times256$ pixels is reduced in $3$ passes with a final
IA size of $64\times64$ pixels with $50$\% overlap in the x and y directions. The time interval
is selected to satisfy the $1/4$ rule \cite{keane1990optimization}. Estimates of the maximum relative error \cite{raffel2018particle} in the PIV measurements is within $1\%$. \\

\begin{figure}[ht]
  \centerline{\includegraphics[clip = true, trim = 0 0 0 0,width= 0.8\textwidth]{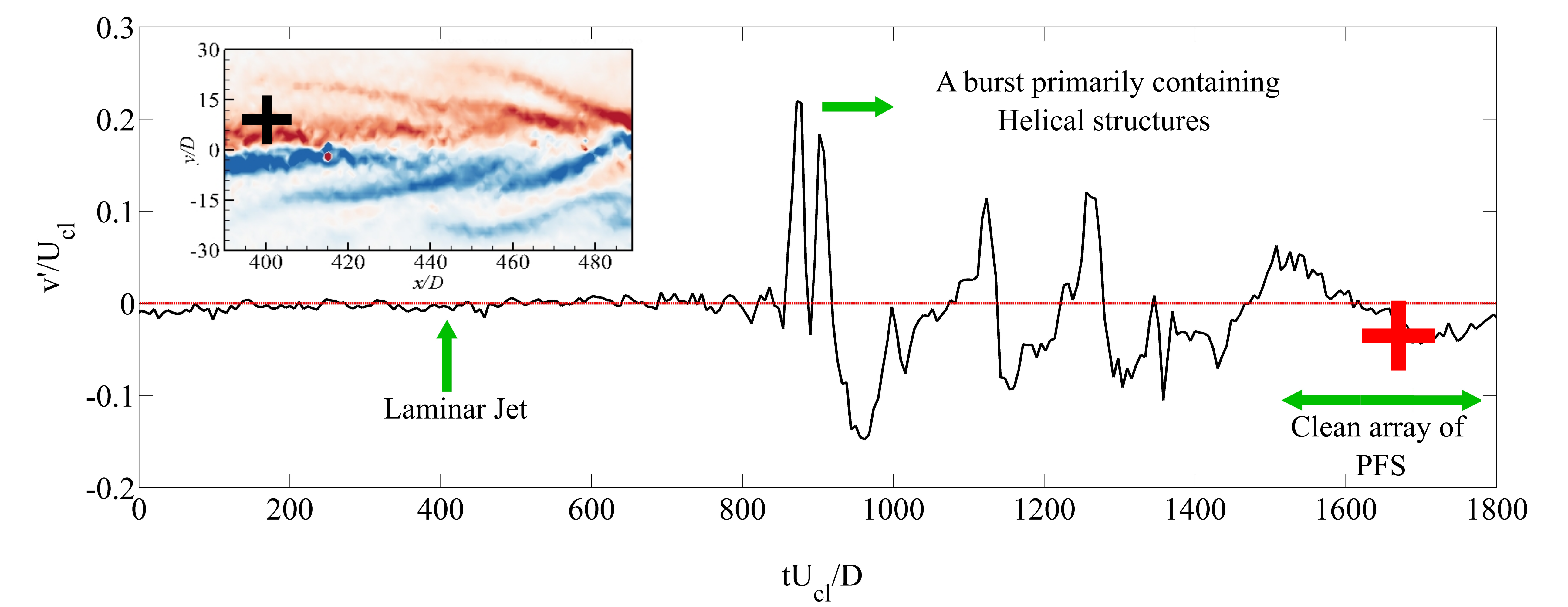}}
 \caption{Time trace of transverse velocity ($v'$) at $x/D=400$ and $y/D=10$ (marked by \textbf{+} sign). The Puff like structures at $tU_{cl}/D = 1676$ (marked by \textcolor{red}{\textbf{+}} sign) is shown in the inset. }
\label{fig:trace}
\end{figure}

 
\begin{figure}[ht]
  \centerline{\includegraphics[clip = true, trim = 0 0 0 0,width= 0.48\textwidth]{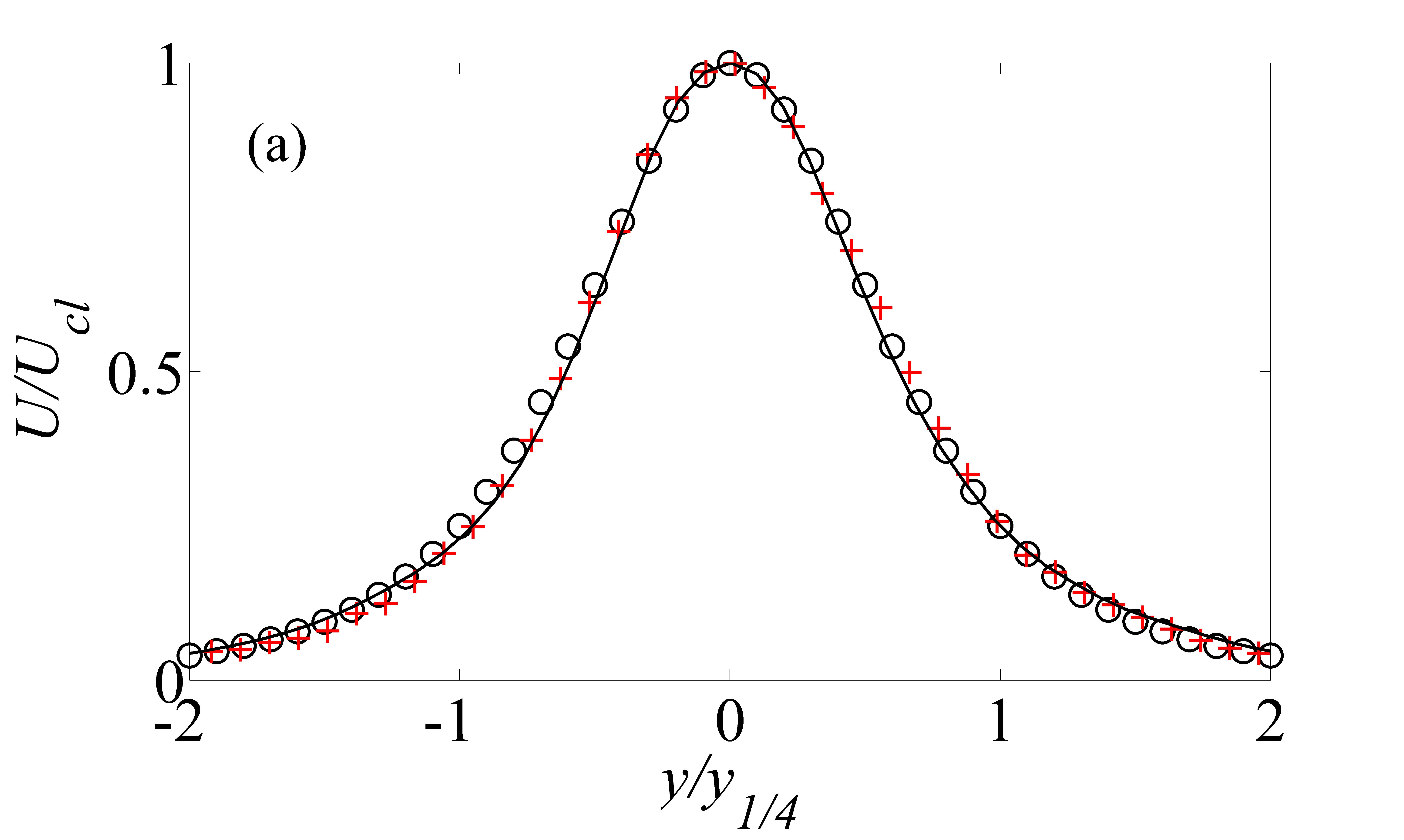}}
  \centerline{
  \includegraphics[clip = true, trim = 0 0 0 0,width= 0.35\textwidth]{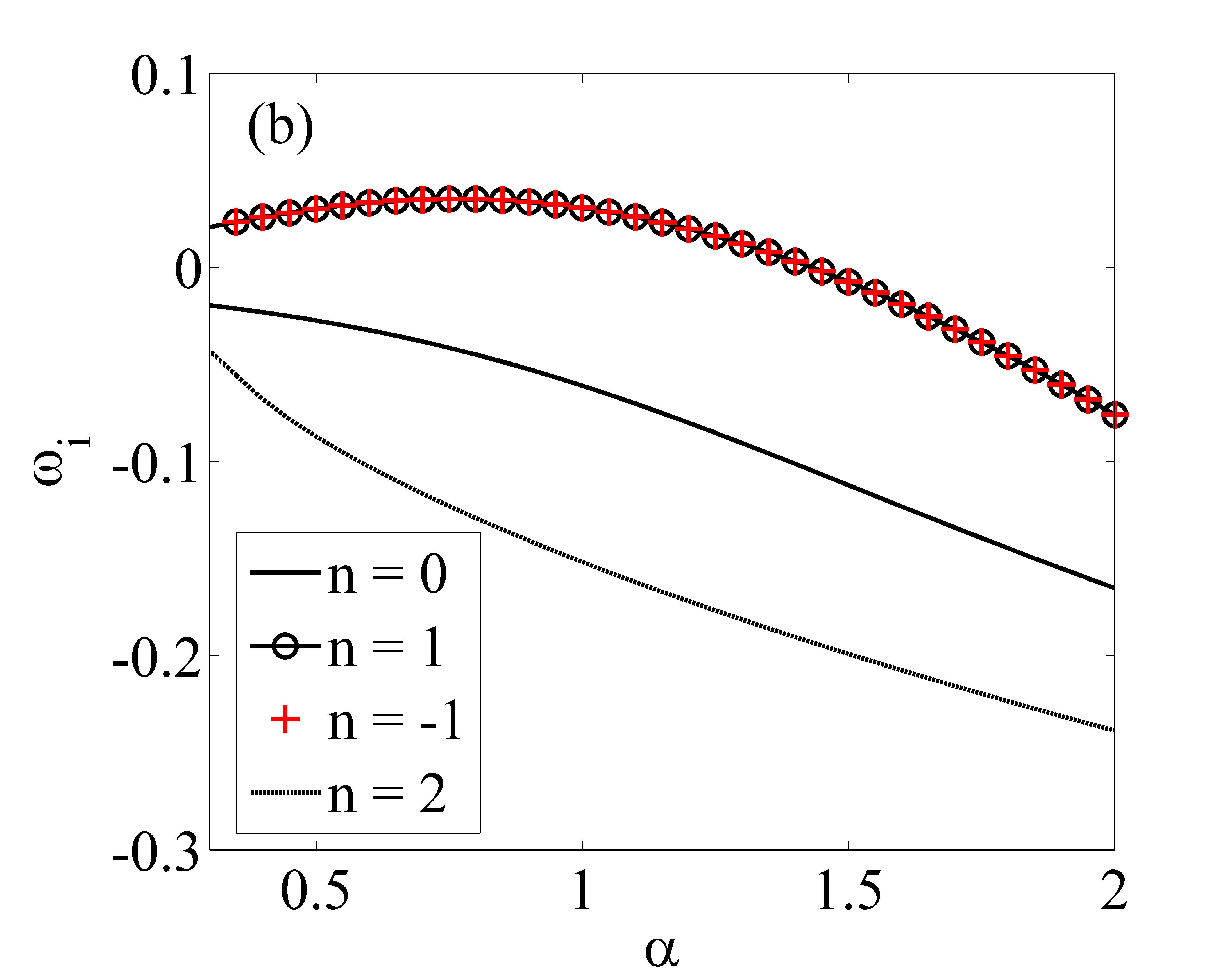}
  \includegraphics[clip = true, trim = 0 0 0 0,width= 0.35\textwidth]{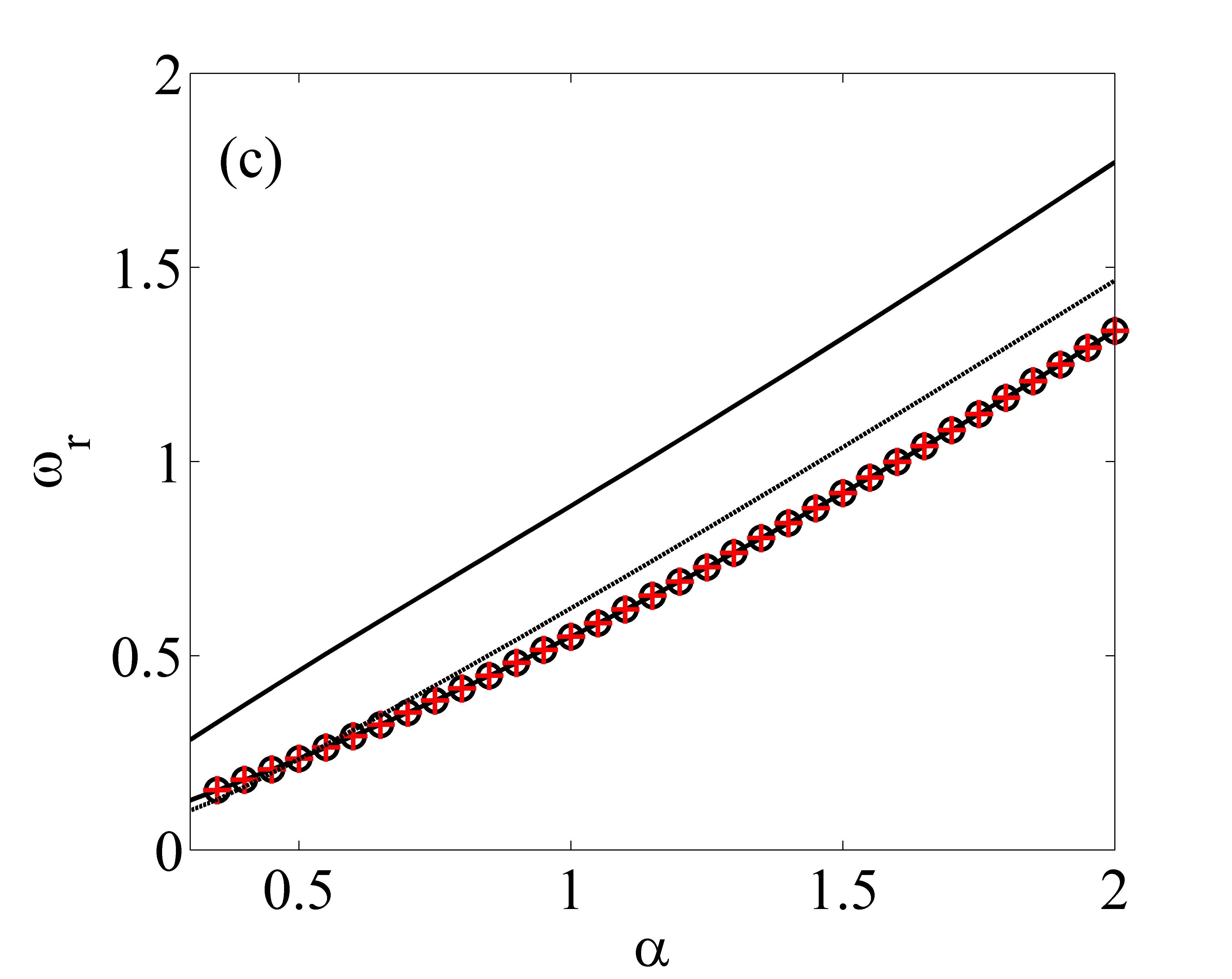}}
 \caption{(a) Nondimensional instantaneous velocity profile ( \RD{+} ) and velocity profiles average over $50 s$ ({ \textemdash } ) at $x/D = 440$ for $Re = 450$, ( o ) analytical self-similar velocity profile given by \cite{landau1959fluid} (b) Growth rate and (c) real part of complex frequency for axisymmetric, helical and double helical modes.}
\label{fig:kf}
\end{figure}

 \noindent \textit{{4. Linear stability analysis}}\\

\textcolor{black}{We observe from the transverse velocity trace in presence of the \emph{Puff-like structures} (PFS) shown in Fig. \ref{fig:trace} here that the transverse velocity fluctuation is large with respect to the jet-centreline velocity and the phenomenon is \emph{non-linear}. Understanding the dynamics using a non-linear model may illustrate mode interaction, phase synchronization and saturation. However, significant insight regarding the modes is obtained from LST; deviations from experimental observations could be explained using a non-linear analysis.} \\

 \noindent \textit{{Base flow}}\\
An instantaneous velocity profile and a velocity profile averaged over $50s$ obtained at $x/D = 440$, are plotted in figure \ref{fig:kf}(a) at $Re=450$ when the jet is laminar and stable. The velocity profiles are non-dimensionalised using local quarter velocity width ($y_{1/4}$) and local jet centerline velocity ($U_{cl}$) \footnote{Global conservation of momentum for the base flow implies that the Reynolds number based on $U_{cl}, y_{1/4}$ is identical to the Reynolds number based on $U_{av}, D$.}. The analytical self similar velocity profile $U = 1/{(1+({r}/{r_o})^2)^2}$ given by \citet{landau1959fluid} is also superimposed in the same plot. All the profiles match closely and we use the self-similar base flow profile for the linear stability analysis. \\

 \noindent \textit{{Methodology}}\\
The linearised Navier-Stokes equations in the cylindrical coordinates for studying the stability of jets have been derived by \citet{morris1976spatial}. We skip stating the equations for brevity and assume that perturbations can be expressed using normal modes (\ref{eq:nma}).
 \begin{equation}
     [u,v,w,p]^T = Real\left\{ [\bar{u}(r),\bar{v}(r),\bar{w}(r),\bar{p}(r)]^T
e^{i(\alpha x + n \phi - \omega t)}\right\} 
\label{eq:nma}
 \end{equation}
where, $\alpha$ is the streamwise wavenumber, $n$ is the azimuthal wavenumber and $\omega$ is the complex frequency. The streamwise, the radial, the azimuthal velocities and the pressure are denoted by $u,v,w,p$. The boundary conditions imposed are that the perturbation decays in the far-field and centreline conditions similar to \citet{batchelor1962analysis}. The governing equations reduce to an eigenvalue problem with real streamwise wavenumber $\alpha$ and complex frequency $\omega$ for the temporal analysis. The derivatives in the radial direction are discretized using the Chebyshev differentiation matrices following \citet{trefethen2000spectral}. The growth rate of the modes corresponding to different azimuthal wavenumbers ($n$) are obtained through LST at $Re = 500$  and are shown in figure \ref{fig:kf}(b). \RevC{The helical modes ($n=\pm 1$) are the only unstable mode in the fully developed region and their growth rates are equal}. While $n=0$ and higher azimuthal modes are stable in the fully developed region \cite{morris1976spatial, batchelor1962analysis}. The real part of complex frequency ($\omega_r$) is shown in figure \ref{fig:kf}(c) for different modes at $Re=500$. Furthermore, Table 1 presents a comparison of the critical Reynolds number of the helical mode obtained from the present computation with several previous studies.

\vspace{1 cm}
\begin{table}[h]
  \begin{center}
\def~{\hphantom{0}}
  \begin{tabular}{lccc}
      \textbf{Authors}  & \textbf{$Re_{cr}$}   &   \textbf{$\alpha$} & \textbf{$\omega_r$} \\[3pt]
      
       \citet{lessen1973stability}   & $37.9$ & $0.3989$ & $0.08$\\
      \citet{morris1976spatial}  & $37.64$ & $0.44$ & $0.1$\\
      \citet{xie2009efficient}   & $37.6828$ & $0.459$ & $0.103$\\
      \citet{kulkarni2007viscous} & $37.68$ & $0.450481$   & $0.104$\\
      Present & $37.6828$ & $0.4505$ & $0.1040$\\
  \end{tabular}
  \caption{Critical Reynolds number of helical mode in fully developed region}
  \label{tab:kg}
  \end{center}
\end{table}

\noindent \textit{{5. Local average velocity calculation}}\\

\begin{figure}[ht]
  \centerline{\includegraphics[clip = true, trim = 0 0 0 0,width= 0.5\textwidth]{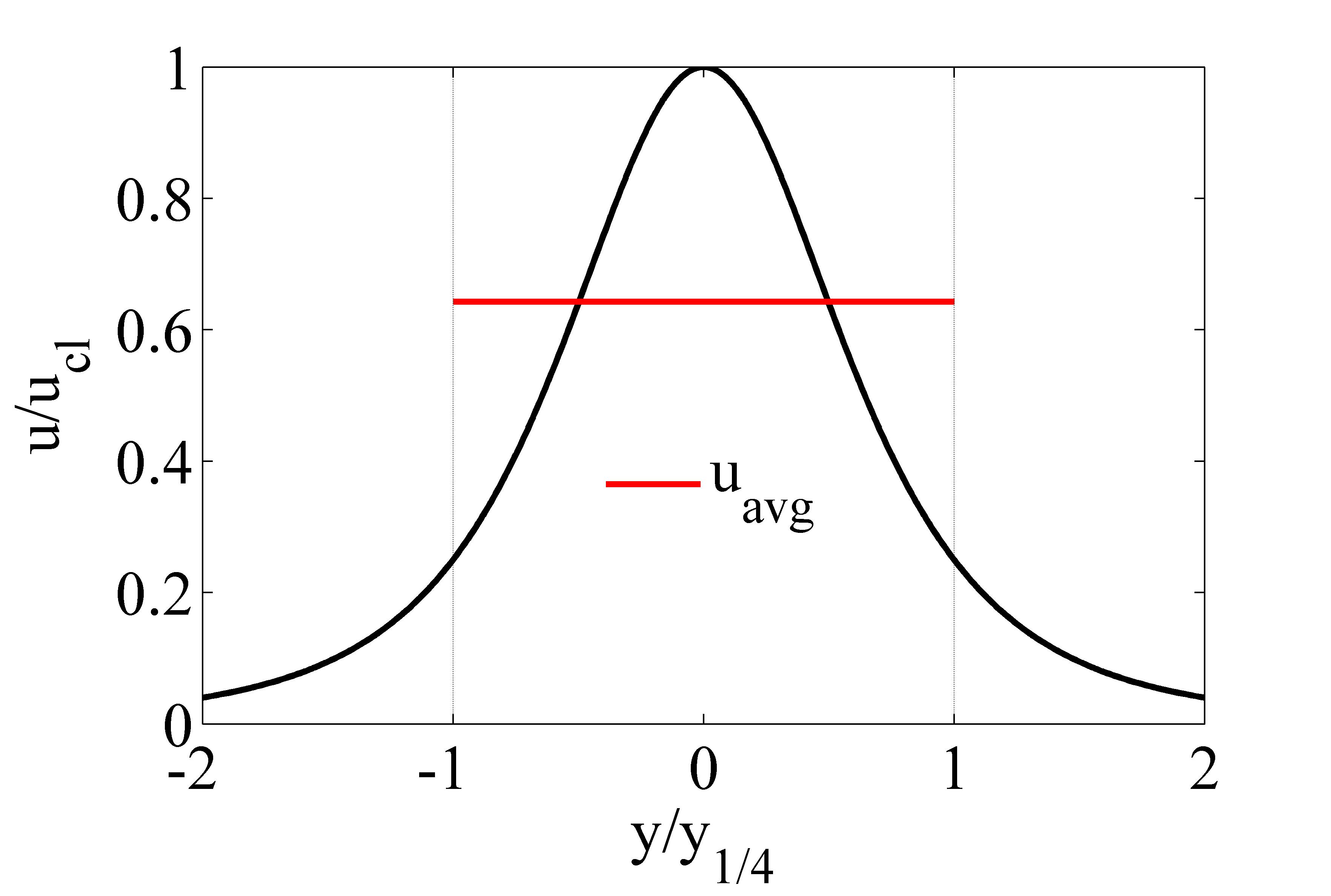}}
 \caption{Average velocity calculation}
\label{fig:avg}
\end{figure}

\noindent The average velocity ($u_{avg}$) is calculated using the local quarter velocity width ($y_{1/4}$). For flow visualization, $u_{avg}$ is calculated using the analytical self-similar velocity profile and local centerline velocity ($U_{cl}$) given by \citet{landau1959fluid}. In PIV experiments, a time averaged velocity profile over $4 $ seconds is obtained at the upstream end of the field of view in presence of puff-like instability (\textit{i.e.}, $x/D =390$). The quarter velocity width is obtained from this profile and the average velocity is calculated within the quarter velocity width as shown in fig. \ref{fig:avg}.\\

\noindent \textit{{6. Calculation of the half angle of the puff like structures from LST and experiments}}\\

\begin{figure}[ht]
  \centerline{\includegraphics[clip = true, trim = 0 0 0 0,width= 0.6\textwidth]{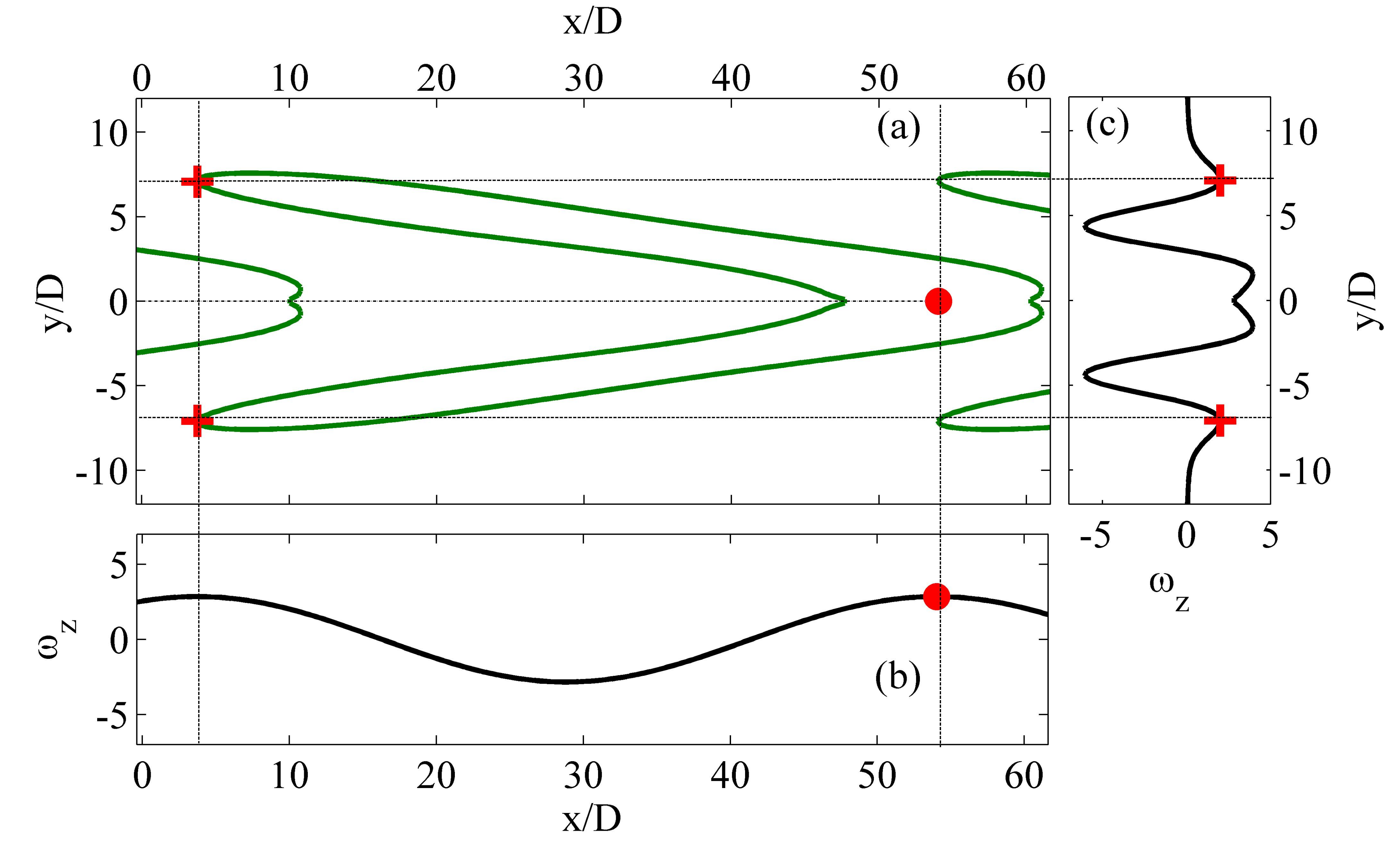}}
 \caption{Visual description of the half angle calculation of a puff-like instability from LST at $\alpha = 1$ and $Re = 550$}
\label{fig:angle}
\end{figure}

\noindent Fig. \ref{fig:angle} describes how we compute the half angles of the puff-like instability. The apex of the instability's leading edge (\textcolor{red}{$\bullet$} sign in fig.s \ref{fig:angle}(a,b)) corresponds to the location of maxima in the eigenfunction $\omega_z(x_m/D,0)$. The trailing edge (\textcolor{red}{$+$} sign in fig.s \ref{fig:angle}(a,c)) corresponds to the radially outermost maxima in the eigenfunction $\omega_z(x_m/D,y_m/D)$. We compute the half-angle of the puff-like instability by joining the leading and trailing edges and measuring the angle with the jet centreline. The green contour shown in fig. \ref{fig:angle}(a) aids in visualizing the instability and corresponds to $\omega_z = 2$ obtained by superimposing $n=\pm 1$ with equal amplitude and phase difference of $\pi$ for $\alpha = 1$ and $Re = 550$.

In the experiments, the upper and lower half angles of the puff like structures were obtained by manually connecting the leading edge and trailing edge of the structures as shown in fig. 4(b) of the manuscript . 

 \newpage
 \noindent \textit{{7. Helical mode frequency obtained from PIV}}\\

\begin{figure}[ht]
  \centerline{\includegraphics[clip = true, trim = 0 0 0 0,width= 0.5\textwidth]{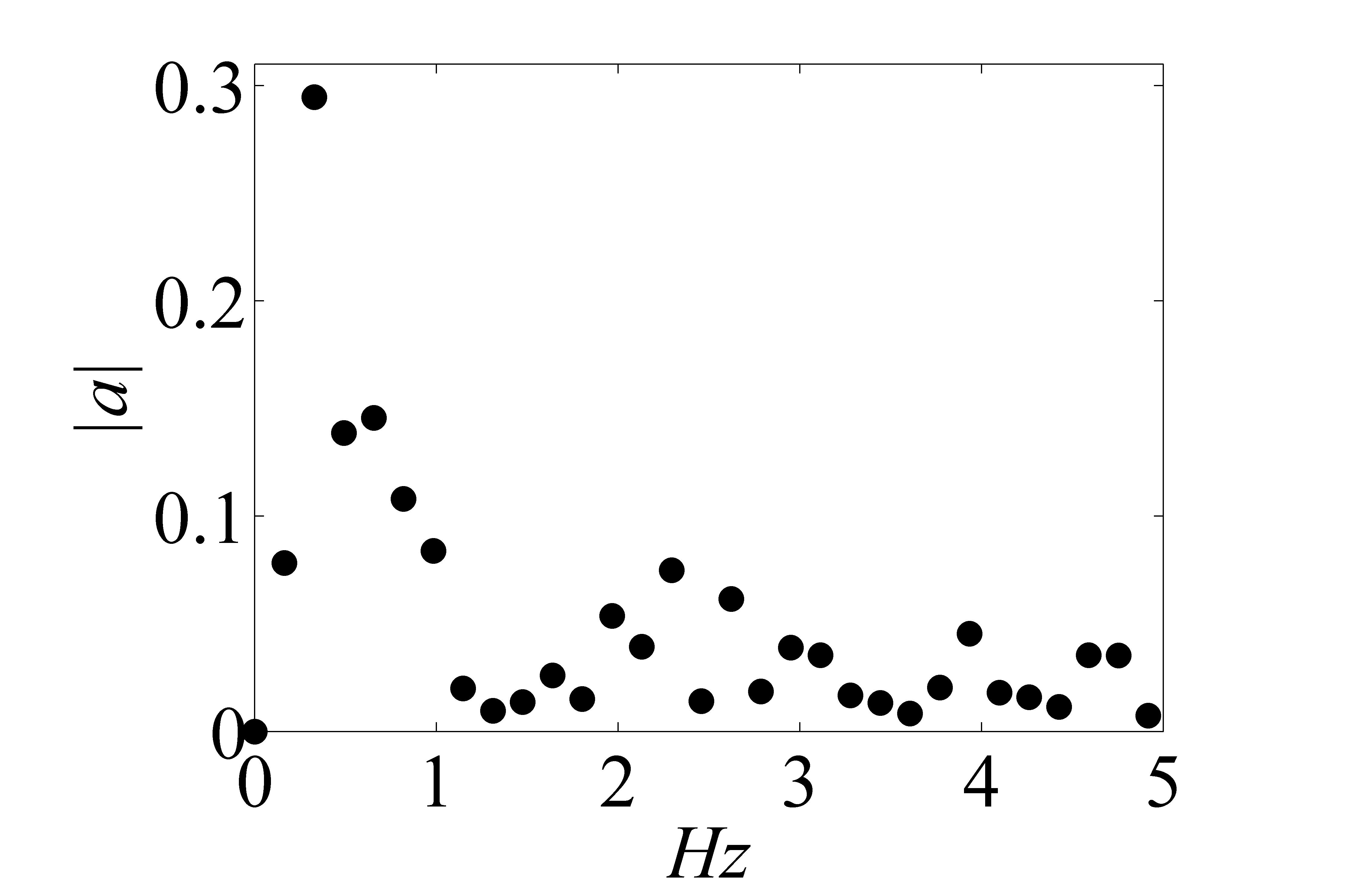}}
 \caption{FFT of time series in Fig. 2(a) of the manuscript}
\label{fig:FFT}
\end{figure}

 \noindent A fast Fourier transform (FFT) of the time series shown in Fig. 2(a) of the manuscript is shown in Fig. \ref{fig:FFT}. The dominant frequency is close to $0.3 Hz$. 

    \begin{figure}[ht]
  \centerline{\includegraphics[clip = true, trim = 0 0 0 0,width= 0.5\textwidth]{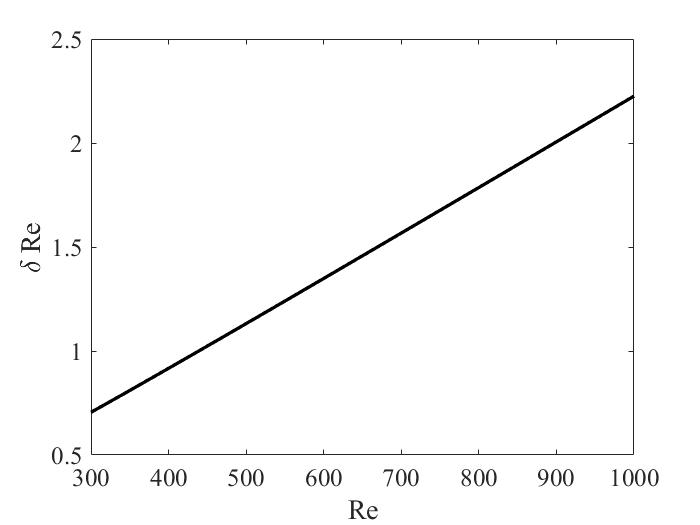}}
 \caption{Uncertainty in $Re$ calculation}
\label{fig:error}
\end{figure}

 \noindent \textit{{8. Uncertainty in Reynolds number calculation}}
 
\RevD{ The Reynolds number was calculated based on the average volume flow rate through the nozzle. The flow rate was measured from the change in the over-head tank's weight placed on a precision weighing machine  (see fig. \ref{fig:SIsetup}) of least count $0.01g$ over a period of $\sim 1 ~\hbox{minute}$. The elapsed time was measured using a stopwatch of least count $0.01s$. As a result, the uncertainty introduced in Reynolds number due to the volume flow rate measurement was negligibly small compared to other sources of errors. For instance, uncertainty in $Re$ measurement was introduced by temperature changes, that were less than $ 0.02^\circ C$ (as evaporation is negligible in the setup) inside the overhead and discharge tanks during the period of an experiment ($1$ minute). Further, the nozzle diameter measurement resolution, $\approx 0.2\%$ of $D$, contributed to the uncertainty in $Re$ calculation. The maximum error in estimated $Re$ is less than $0.25\%$ of $Re$ (see fig. \ref{fig:error}).}

\bibliography{bibliography}

\begin{thebibliography}{51}%
\makeatletter
\providecommand \@ifxundefined [1]{%
 \@ifx{#1\undefined}
}%
\providecommand \@ifnum [1]{%
 \ifnum #1\expandafter \@firstoftwo
 \else \expandafter \@secondoftwo
 \fi
}%
\providecommand \@ifx [1]{%
 \ifx #1\expandafter \@firstoftwo
 \else \expandafter \@secondoftwo
 \fi
}%
\providecommand \natexlab [1]{#1}%
\providecommand \enquote  [1]{``#1''}%
\providecommand \bibnamefont  [1]{#1}%
\providecommand \bibfnamefont [1]{#1}%
\providecommand \citenamefont [1]{#1}%
\providecommand \href@noop [0]{\@secondoftwo}%
\providecommand \href [0]{\begingroup \@sanitize@url \@href}%
\providecommand \@href[1]{\@@startlink{#1}\@@href}%
\providecommand \@@href[1]{\endgroup#1\@@endlink}%
\providecommand \@sanitize@url [0]{\catcode `\\12\catcode `\$12\catcode
  `\&12\catcode `\#12\catcode `\^12\catcode `\_12\catcode `\%12\relax}%
\providecommand \@@startlink[1]{}%
\providecommand \@@endlink[0]{}%
\providecommand \url  [0]{\begingroup\@sanitize@url \@url }%
\providecommand \@url [1]{\endgroup\@href {#1}{\urlprefix }}%
\providecommand \urlprefix  [0]{URL }%
\providecommand \Eprint [0]{\href }%
\providecommand \doibase [0]{http://dx.doi.org/}%
\providecommand \selectlanguage [0]{\@gobble}%
\providecommand \bibinfo  [0]{\@secondoftwo}%
\providecommand \bibfield  [0]{\@secondoftwo}%
\providecommand \translation [1]{[#1]}%
\providecommand \BibitemOpen [0]{}%
\providecommand \bibitemStop [0]{}%
\providecommand \bibitemNoStop [0]{.\EOS\space}%
\providecommand \EOS [0]{\spacefactor3000\relax}%
\providecommand \BibitemShut  [1]{\csname bibitem#1\endcsname}%
\let\auto@bib@innerbib\@empty
\bibitem [{\citenamefont {Meier}\ \emph {et~al.}(2001)\citenamefont {Meier},
  \citenamefont {Koide},\ and\ \citenamefont
  {Uchida}}]{meier2001magnetohydrodynamic}%
  \BibitemOpen
  \bibfield  {author} {\bibinfo {author} {\bibfnamefont {D.~L.}\ \bibnamefont
  {Meier}}, \bibinfo {author} {\bibfnamefont {S.}~\bibnamefont {Koide}}, \ and\
  \bibinfo {author} {\bibfnamefont {Y.}~\bibnamefont {Uchida}},\ }\href@noop {}
  {\bibfield  {journal} {\bibinfo  {journal} {Science}\ }\textbf {\bibinfo
  {volume} {291}},\ \bibinfo {pages} {84} (\bibinfo {year} {2001})}\BibitemShut
  {NoStop}%
\bibitem [{\citenamefont {Joseph}\ \emph {et~al.}(2003)\citenamefont {Joseph},
  \citenamefont {Mahalov}, \citenamefont {Nicolaenko},\ and\ \citenamefont
  {Tse}}]{joseph2003high}%
  \BibitemOpen
  \bibfield  {author} {\bibinfo {author} {\bibfnamefont {B.}~\bibnamefont
  {Joseph}}, \bibinfo {author} {\bibfnamefont {A.}~\bibnamefont {Mahalov}},
  \bibinfo {author} {\bibfnamefont {B.}~\bibnamefont {Nicolaenko}}, \ and\
  \bibinfo {author} {\bibfnamefont {K.~L.}\ \bibnamefont {Tse}},\ }\href@noop
  {} {\bibfield  {journal} {\bibinfo  {journal} {Geophys. Res. Lett.}\ }\textbf
  {\bibinfo {volume} {30}} (\bibinfo {year} {2003})}\BibitemShut {NoStop}%
\bibitem [{\citenamefont {Tang}\ \emph {et~al.}(2013)\citenamefont {Tang},
  \citenamefont {Nicolle}, \citenamefont {Klettner}, \citenamefont {Pantelic},
  \citenamefont {Wang}, \citenamefont {Suhaimi}, \citenamefont {Tan},
  \citenamefont {Ong}, \citenamefont {Su}, \citenamefont {Sekhar} \emph
  {et~al.}}]{tang2013airflow}%
  \BibitemOpen
  \bibfield  {author} {\bibinfo {author} {\bibfnamefont {J.~W.}\ \bibnamefont
  {Tang}}, \bibinfo {author} {\bibfnamefont {A.~D.}\ \bibnamefont {Nicolle}},
  \bibinfo {author} {\bibfnamefont {C.~A.}\ \bibnamefont {Klettner}}, \bibinfo
  {author} {\bibfnamefont {J.}~\bibnamefont {Pantelic}}, \bibinfo {author}
  {\bibfnamefont {L.}~\bibnamefont {Wang}}, \bibinfo {author} {\bibfnamefont
  {A.~B.}\ \bibnamefont {Suhaimi}}, \bibinfo {author} {\bibfnamefont
  {A.~Y.~L.}\ \bibnamefont {Tan}}, \bibinfo {author} {\bibfnamefont {G.~W.~X.}\
  \bibnamefont {Ong}}, \bibinfo {author} {\bibfnamefont {R.}~\bibnamefont
  {Su}}, \bibinfo {author} {\bibfnamefont {C.}~\bibnamefont {Sekhar}},  \emph
  {et~al.},\ }\href@noop {} {\bibfield  {journal} {\bibinfo  {journal} {PLoS
  One}\ }\textbf {\bibinfo {volume} {8}},\ \bibinfo {pages} {e59970} (\bibinfo
  {year} {2013})}\BibitemShut {NoStop}%
\bibitem [{\citenamefont {Dabiri}\ \emph {et~al.}(2010)\citenamefont {Dabiri},
  \citenamefont {Colin}, \citenamefont {Katija},\ and\ \citenamefont
  {Costello}}]{dabiri2010wake}%
  \BibitemOpen
  \bibfield  {author} {\bibinfo {author} {\bibfnamefont {J.~O.}\ \bibnamefont
  {Dabiri}}, \bibinfo {author} {\bibfnamefont {S.~P.}\ \bibnamefont {Colin}},
  \bibinfo {author} {\bibfnamefont {K.}~\bibnamefont {Katija}}, \ and\ \bibinfo
  {author} {\bibfnamefont {J.~H.}\ \bibnamefont {Costello}},\ }\href@noop {}
  {\bibfield  {journal} {\bibinfo  {journal} {J. Exp. Biol.}\ }\textbf
  {\bibinfo {volume} {213}},\ \bibinfo {pages} {1217} (\bibinfo {year}
  {2010})}\BibitemShut {NoStop}%
\bibitem [{\citenamefont {Nathan}\ \emph {et~al.}(2006)\citenamefont {Nathan},
  \citenamefont {Mi}, \citenamefont {Alwahabi}, \citenamefont {Newbold},\ and\
  \citenamefont {Nobes}}]{nathan2006impacts}%
  \BibitemOpen
  \bibfield  {author} {\bibinfo {author} {\bibfnamefont {G.~J.}\ \bibnamefont
  {Nathan}}, \bibinfo {author} {\bibfnamefont {J.}~\bibnamefont {Mi}}, \bibinfo
  {author} {\bibfnamefont {Z.~T.}\ \bibnamefont {Alwahabi}}, \bibinfo {author}
  {\bibfnamefont {G.~J.~R.}\ \bibnamefont {Newbold}}, \ and\ \bibinfo {author}
  {\bibfnamefont {D.~S.}\ \bibnamefont {Nobes}},\ }\href@noop {} {\bibfield
  {journal} {\bibinfo  {journal} {Prog. Energy Combust. Sci.}\ }\textbf
  {\bibinfo {volume} {32}},\ \bibinfo {pages} {496} (\bibinfo {year}
  {2006})}\BibitemShut {NoStop}%
\bibitem [{\citenamefont {Levchenko}\ \emph {et~al.}(2018)\citenamefont
  {Levchenko}, \citenamefont {Xu}, \citenamefont {Teel}, \citenamefont
  {Mariotti}, \citenamefont {Walker},\ and\ \citenamefont
  {Keidar}}]{levchenko2018recent}%
  \BibitemOpen
  \bibfield  {author} {\bibinfo {author} {\bibfnamefont {I.}~\bibnamefont
  {Levchenko}}, \bibinfo {author} {\bibfnamefont {S.}~\bibnamefont {Xu}},
  \bibinfo {author} {\bibfnamefont {G.}~\bibnamefont {Teel}}, \bibinfo {author}
  {\bibfnamefont {D.}~\bibnamefont {Mariotti}}, \bibinfo {author}
  {\bibfnamefont {M.~L.~R.}\ \bibnamefont {Walker}}, \ and\ \bibinfo {author}
  {\bibfnamefont {M.}~\bibnamefont {Keidar}},\ }\href@noop {} {\bibfield
  {journal} {\bibinfo  {journal} {Nat. Commun.}\ }\textbf {\bibinfo {volume}
  {9}},\ \bibinfo {pages} {1} (\bibinfo {year} {2018})}\BibitemShut {NoStop}%
\bibitem [{\citenamefont {Schramm-Baxter}\ and\ \citenamefont
  {Mitragotri}(2004)}]{schramm2004needle}%
  \BibitemOpen
  \bibfield  {author} {\bibinfo {author} {\bibfnamefont {J.}~\bibnamefont
  {Schramm-Baxter}}\ and\ \bibinfo {author} {\bibfnamefont {S.}~\bibnamefont
  {Mitragotri}},\ }\href@noop {} {\bibfield  {journal} {\bibinfo  {journal} {J.
  Control. Release}\ }\textbf {\bibinfo {volume} {97}},\ \bibinfo {pages} {527}
  (\bibinfo {year} {2004})}\BibitemShut {NoStop}%
\bibitem [{\citenamefont {Bejan}\ \emph {et~al.}(2014)\citenamefont {Bejan},
  \citenamefont {Ziaei},\ and\ \citenamefont {Lorente}}]{bejan2014evolution}%
  \BibitemOpen
  \bibfield  {author} {\bibinfo {author} {\bibfnamefont {A.}~\bibnamefont
  {Bejan}}, \bibinfo {author} {\bibfnamefont {S.}~\bibnamefont {Ziaei}}, \ and\
  \bibinfo {author} {\bibfnamefont {S.}~\bibnamefont {Lorente}},\ }\href@noop
  {} {\bibfield  {journal} {\bibinfo  {journal} {Sci. Rep.}\ }\textbf {\bibinfo
  {volume} {4}},\ \bibinfo {pages} {4730} (\bibinfo {year} {2014})}\BibitemShut
  {NoStop}%
\bibitem [{\citenamefont {Cabanes}\ \emph {et~al.}(2017)\citenamefont
  {Cabanes}, \citenamefont {Aurnou}, \citenamefont {Favier},\ and\
  \citenamefont {Le~Bars}}]{cabanes2017laboratory}%
  \BibitemOpen
  \bibfield  {author} {\bibinfo {author} {\bibfnamefont {S.}~\bibnamefont
  {Cabanes}}, \bibinfo {author} {\bibfnamefont {J.}~\bibnamefont {Aurnou}},
  \bibinfo {author} {\bibfnamefont {B.}~\bibnamefont {Favier}}, \ and\ \bibinfo
  {author} {\bibfnamefont {M.}~\bibnamefont {Le~Bars}},\ }\href@noop {}
  {\bibfield  {journal} {\bibinfo  {journal} {Nat. Phys.}\ }\textbf {\bibinfo
  {volume} {13}},\ \bibinfo {pages} {387} (\bibinfo {year} {2017})}\BibitemShut
  {NoStop}%
\bibitem [{\citenamefont {Torres-Alb{\`a}}(2020)}]{torres2020stars}%
  \BibitemOpen
  \bibfield  {author} {\bibinfo {author} {\bibfnamefont {N.}~\bibnamefont
  {Torres-Alb{\`a}}},\ }\href@noop {} {\bibfield  {journal} {\bibinfo
  {journal} {Nat. Astron.}\ }\textbf {\bibinfo {volume} {4}},\ \bibinfo {pages}
  {448} (\bibinfo {year} {2020})}\BibitemShut {NoStop}%
\bibitem [{\citenamefont {Underwood}\ \emph {et~al.}(2019)\citenamefont
  {Underwood}, \citenamefont {Loebner}, \citenamefont {Miller},\ and\
  \citenamefont {Cappelli}}]{underwood2019dynamic}%
  \BibitemOpen
  \bibfield  {author} {\bibinfo {author} {\bibfnamefont {T.~C.}\ \bibnamefont
  {Underwood}}, \bibinfo {author} {\bibfnamefont {K.~T.~K.}\ \bibnamefont
  {Loebner}}, \bibinfo {author} {\bibfnamefont {V.~A.}\ \bibnamefont {Miller}},
  \ and\ \bibinfo {author} {\bibfnamefont {M.~A.}\ \bibnamefont {Cappelli}},\
  }\href@noop {} {\bibfield  {journal} {\bibinfo  {journal} {Sci. Rep.}\
  }\textbf {\bibinfo {volume} {9}},\ \bibinfo {pages} {1} (\bibinfo {year}
  {2019})}\BibitemShut {NoStop}%
\bibitem [{\citenamefont {Mossa}\ and\ \citenamefont
  {De~Serio}(2016)}]{mossa2016rethinking}%
  \BibitemOpen
  \bibfield  {author} {\bibinfo {author} {\bibfnamefont {M.}~\bibnamefont
  {Mossa}}\ and\ \bibinfo {author} {\bibfnamefont {F.}~\bibnamefont
  {De~Serio}},\ }\href@noop {} {\bibfield  {journal} {\bibinfo  {journal} {Sci.
  Rep.}\ }\textbf {\bibinfo {volume} {6}},\ \bibinfo {pages} {1} (\bibinfo
  {year} {2016})}\BibitemShut {NoStop}%
\bibitem [{\citenamefont {Mitsudera}\ \emph {et~al.}(2018)\citenamefont
  {Mitsudera}, \citenamefont {Miyama}, \citenamefont {Nishigaki}, \citenamefont
  {Nakanowatari}, \citenamefont {Nishikawa}, \citenamefont {Nakamura},
  \citenamefont {Wagawa}, \citenamefont {Furue}, \citenamefont {Fujii},\ and\
  \citenamefont {Ito}}]{mitsudera2018low}%
  \BibitemOpen
  \bibfield  {author} {\bibinfo {author} {\bibfnamefont {H.}~\bibnamefont
  {Mitsudera}}, \bibinfo {author} {\bibfnamefont {T.}~\bibnamefont {Miyama}},
  \bibinfo {author} {\bibfnamefont {H.}~\bibnamefont {Nishigaki}}, \bibinfo
  {author} {\bibfnamefont {T.}~\bibnamefont {Nakanowatari}}, \bibinfo {author}
  {\bibfnamefont {H.}~\bibnamefont {Nishikawa}}, \bibinfo {author}
  {\bibfnamefont {T.}~\bibnamefont {Nakamura}}, \bibinfo {author}
  {\bibfnamefont {T.}~\bibnamefont {Wagawa}}, \bibinfo {author} {\bibfnamefont
  {R.}~\bibnamefont {Furue}}, \bibinfo {author} {\bibfnamefont
  {Y.}~\bibnamefont {Fujii}}, \ and\ \bibinfo {author} {\bibfnamefont
  {S.}~\bibnamefont {Ito}},\ }\href@noop {} {\bibfield  {journal} {\bibinfo
  {journal} {Nat. Commun.}\ }\textbf {\bibinfo {volume} {9}},\ \bibinfo {pages}
  {1} (\bibinfo {year} {2018})}\BibitemShut {NoStop}%
\bibitem [{\citenamefont {Reynolds}(1962)}]{reynolds1962observations}%
  \BibitemOpen
  \bibfield  {author} {\bibinfo {author} {\bibfnamefont {A.~J.}\ \bibnamefont
  {Reynolds}},\ }\href@noop {} {\bibfield  {journal} {\bibinfo  {journal} {J.
  Fluid Mech.}\ }\textbf {\bibinfo {volume} {14}},\ \bibinfo {pages} {552}
  (\bibinfo {year} {1962})}\BibitemShut {NoStop}%
\bibitem [{\citenamefont {Viilu}(1962)}]{viilu1962experimental}%
  \BibitemOpen
  \bibfield  {author} {\bibinfo {author} {\bibfnamefont {A.}~\bibnamefont
  {Viilu}},\ }\href@noop {} {\bibfield  {journal} {\bibinfo  {journal} {J.
  Appl. Mech.}\ }\textbf {\bibinfo {volume} {29}},\ \bibinfo {pages} {506}
  (\bibinfo {year} {1962})}\BibitemShut {NoStop}%
\bibitem [{\citenamefont {O’Neill}\ \emph {et~al.}(2004)\citenamefont
  {O’Neill}, \citenamefont {Soria},\ and\ \citenamefont
  {Honnery}}]{o2004stability}%
  \BibitemOpen
  \bibfield  {author} {\bibinfo {author} {\bibfnamefont {P.}~\bibnamefont
  {O’Neill}}, \bibinfo {author} {\bibfnamefont {J.}~\bibnamefont {Soria}}, \
  and\ \bibinfo {author} {\bibfnamefont {D.}~\bibnamefont {Honnery}},\
  }\href@noop {} {\bibfield  {journal} {\bibinfo  {journal} {Exp. Fluids}\
  }\textbf {\bibinfo {volume} {36}},\ \bibinfo {pages} {473} (\bibinfo {year}
  {2004})}\BibitemShut {NoStop}%
\bibitem [{\citenamefont {Yule}(1978)}]{yule1978large}%
  \BibitemOpen
  \bibfield  {author} {\bibinfo {author} {\bibfnamefont {A.}~\bibnamefont
  {Yule}},\ }\href@noop {} {\bibfield  {journal} {\bibinfo  {journal} {Journal
  of Fluid Mechanics}\ }\textbf {\bibinfo {volume} {89}},\ \bibinfo {pages}
  {413} (\bibinfo {year} {1978})}\BibitemShut {NoStop}%
\bibitem [{\citenamefont {Verzicco}\ and\ \citenamefont
  {Orlandi}(1994)}]{verzicco1994direct}%
  \BibitemOpen
  \bibfield  {author} {\bibinfo {author} {\bibfnamefont {R.}~\bibnamefont
  {Verzicco}}\ and\ \bibinfo {author} {\bibfnamefont {P.}~\bibnamefont
  {Orlandi}},\ }\href@noop {} {\bibfield  {journal} {\bibinfo  {journal} {Phys.
  Fluids}\ }\textbf {\bibinfo {volume} {6}},\ \bibinfo {pages} {751} (\bibinfo
  {year} {1994})}\BibitemShut {NoStop}%
\bibitem [{\citenamefont {Mattingly}\ and\ \citenamefont
  {Chang}(1974)}]{mattingly1974unstable}%
  \BibitemOpen
  \bibfield  {author} {\bibinfo {author} {\bibfnamefont {G.~E.}\ \bibnamefont
  {Mattingly}}\ and\ \bibinfo {author} {\bibfnamefont {C.~C.}\ \bibnamefont
  {Chang}},\ }\href@noop {} {\bibfield  {journal} {\bibinfo  {journal} {J.
  Fluid Mech.}\ }\textbf {\bibinfo {volume} {65}},\ \bibinfo {pages} {541}
  (\bibinfo {year} {1974})}\BibitemShut {NoStop}%
\bibitem [{\citenamefont {Mullin}(2011)}]{mullin2011experimental}%
  \BibitemOpen
  \bibfield  {author} {\bibinfo {author} {\bibfnamefont {T.}~\bibnamefont
  {Mullin}},\ }\href@noop {} {\bibfield  {journal} {\bibinfo  {journal} {Annu.
  Rev. Fluid Mech.}\ }\textbf {\bibinfo {volume} {43}},\ \bibinfo {pages} {1}
  (\bibinfo {year} {2011})}\BibitemShut {NoStop}%
\bibitem [{\citenamefont {Hof}\ \emph {et~al.}(2010)\citenamefont {Hof},
  \citenamefont {De~Lozar}, \citenamefont {Avila}, \citenamefont {Tu},\ and\
  \citenamefont {Schneider}}]{hof2010eliminating}%
  \BibitemOpen
  \bibfield  {author} {\bibinfo {author} {\bibfnamefont {B.}~\bibnamefont
  {Hof}}, \bibinfo {author} {\bibfnamefont {A.}~\bibnamefont {De~Lozar}},
  \bibinfo {author} {\bibfnamefont {M.}~\bibnamefont {Avila}}, \bibinfo
  {author} {\bibfnamefont {X.}~\bibnamefont {Tu}}, \ and\ \bibinfo {author}
  {\bibfnamefont {T.~M.}\ \bibnamefont {Schneider}},\ }\href@noop {} {\bibfield
   {journal} {\bibinfo  {journal} {Science}\ }\textbf {\bibinfo {volume}
  {327}},\ \bibinfo {pages} {1491} (\bibinfo {year} {2010})}\BibitemShut
  {NoStop}%
\bibitem [{\citenamefont {Tsukahara}\ \emph {et~al.}(2005)\citenamefont
  {Tsukahara}, \citenamefont {Seki}, \citenamefont {Kawamura},\ and\
  \citenamefont {Tochio}}]{tsukahara2005dns}%
  \BibitemOpen
  \bibfield  {author} {\bibinfo {author} {\bibfnamefont {T.}~\bibnamefont
  {Tsukahara}}, \bibinfo {author} {\bibfnamefont {Y.}~\bibnamefont {Seki}},
  \bibinfo {author} {\bibfnamefont {H.}~\bibnamefont {Kawamura}}, \ and\
  \bibinfo {author} {\bibfnamefont {D.}~\bibnamefont {Tochio}},\ }in\
  \href@noop {} {\emph {\bibinfo {booktitle} {$IV^{th}$ TSFP}}}\ (\bibinfo
  {organization} {Begel House Inc.},\ \bibinfo {year} {2005})\BibitemShut
  {NoStop}%
\bibitem [{\citenamefont {Vinod}\ and\ \citenamefont
  {Govindarajan}(2004)}]{vinod2004pattern}%
  \BibitemOpen
  \bibfield  {author} {\bibinfo {author} {\bibfnamefont {N.}~\bibnamefont
  {Vinod}}\ and\ \bibinfo {author} {\bibfnamefont {R.}~\bibnamefont
  {Govindarajan}},\ }\href@noop {} {\bibfield  {journal} {\bibinfo  {journal}
  {Phys. Rev. Lett.}\ }\textbf {\bibinfo {volume} {93}},\ \bibinfo {pages}
  {114501} (\bibinfo {year} {2004})}\BibitemShut {NoStop}%
\bibitem [{\citenamefont {Chantry}\ \emph {et~al.}(2016)\citenamefont
  {Chantry}, \citenamefont {Tuckerman},\ and\ \citenamefont
  {Barkley}}]{chantry2016turbulent}%
  \BibitemOpen
  \bibfield  {author} {\bibinfo {author} {\bibfnamefont {M.}~\bibnamefont
  {Chantry}}, \bibinfo {author} {\bibfnamefont {L.~S.}\ \bibnamefont
  {Tuckerman}}, \ and\ \bibinfo {author} {\bibfnamefont {D.}~\bibnamefont
  {Barkley}},\ }\href@noop {} {\bibfield  {journal} {\bibinfo  {journal} {J.
  Fluid Mech.}\ }\textbf {\bibinfo {volume} {791}} (\bibinfo {year}
  {2016})}\BibitemShut {NoStop}%
\bibitem [{\citenamefont {Tuckerman}\ \emph {et~al.}(2020)\citenamefont
  {Tuckerman}, \citenamefont {Chantry},\ and\ \citenamefont
  {Barkley}}]{tuckerman2020patterns}%
  \BibitemOpen
  \bibfield  {author} {\bibinfo {author} {\bibfnamefont {L.~S.}\ \bibnamefont
  {Tuckerman}}, \bibinfo {author} {\bibfnamefont {M.}~\bibnamefont {Chantry}},
  \ and\ \bibinfo {author} {\bibfnamefont {D.}~\bibnamefont {Barkley}},\
  }\href@noop {} {\bibfield  {journal} {\bibinfo  {journal} {Annu. Rev. Fluid
  Mech.}\ }\textbf {\bibinfo {volume} {52}} (\bibinfo {year}
  {2020})}\BibitemShut {NoStop}%
\bibitem [{\citenamefont {Landau}(1944)}]{landau1944problem}%
  \BibitemOpen
  \bibfield  {author} {\bibinfo {author} {\bibfnamefont {L.~D.}\ \bibnamefont
  {Landau}},\ }in\ \href@noop {} {\emph {\bibinfo {booktitle} {Dokl. Akad. Nauk
  USSR}}},\ Vol.~\bibinfo {volume} {44}\ (\bibinfo {year} {1944})\ p.\ \bibinfo
  {pages} {311}\BibitemShut {NoStop}%
\bibitem [{\citenamefont {Pomeau}(2016)}]{pomeau2016long}%
  \BibitemOpen
  \bibfield  {author} {\bibinfo {author} {\bibfnamefont {Y.}~\bibnamefont
  {Pomeau}},\ }\href@noop {} {\bibfield  {journal} {\bibinfo  {journal} {Nat.
  Phys.}\ }\textbf {\bibinfo {volume} {12}},\ \bibinfo {pages} {198} (\bibinfo
  {year} {2016})}\BibitemShut {NoStop}%
\bibitem [{\citenamefont {Govindarajan}\ and\ \citenamefont
  {Narasimha}(2000)}]{govindarajan_narasimha_2000}%
  \BibitemOpen
  \bibfield  {author} {\bibinfo {author} {\bibfnamefont {R.}~\bibnamefont
  {Govindarajan}}\ and\ \bibinfo {author} {\bibfnamefont {R.}~\bibnamefont
  {Narasimha}},\ }\href {\doibase 10.1017/S002211200000104X} {\bibfield
  {journal} {\bibinfo  {journal} {J. Fluid Mech.}\ }\textbf {\bibinfo {volume}
  {418}},\ \bibinfo {pages} {77} (\bibinfo {year} {2000})}\BibitemShut
  {NoStop}%
\bibitem [{\citenamefont {Batchelor}\ and\ \citenamefont
  {Gill}(1962)}]{batchelor1962analysis}%
  \BibitemOpen
  \bibfield  {author} {\bibinfo {author} {\bibfnamefont {G.~K.}\ \bibnamefont
  {Batchelor}}\ and\ \bibinfo {author} {\bibfnamefont {A.~E.}\ \bibnamefont
  {Gill}},\ }\href@noop {} {\bibfield  {journal} {\bibinfo  {journal} {J. Fluid
  Mech.}\ }\textbf {\bibinfo {volume} {14}},\ \bibinfo {pages} {529} (\bibinfo
  {year} {1962})}\BibitemShut {NoStop}%
\bibitem [{\citenamefont {Lessen}\ and\ \citenamefont
  {Singh}(1973)}]{lessen1973stability}%
  \BibitemOpen
  \bibfield  {author} {\bibinfo {author} {\bibfnamefont {M.}~\bibnamefont
  {Lessen}}\ and\ \bibinfo {author} {\bibfnamefont {P.~J.}\ \bibnamefont
  {Singh}},\ }\href@noop {} {\bibfield  {journal} {\bibinfo  {journal} {Journal
  of Fluid Mechanics}\ }\textbf {\bibinfo {volume} {60}},\ \bibinfo {pages}
  {433} (\bibinfo {year} {1973})}\BibitemShut {NoStop}%
\bibitem [{\citenamefont {Morris}(1976)}]{morris1976spatial}%
  \BibitemOpen
  \bibfield  {author} {\bibinfo {author} {\bibfnamefont {P.~J.}\ \bibnamefont
  {Morris}},\ }\href@noop {} {\bibfield  {journal} {\bibinfo  {journal}
  {Journal of Fluid Mechanics}\ }\textbf {\bibinfo {volume} {77}},\ \bibinfo
  {pages} {511} (\bibinfo {year} {1976})}\BibitemShut {NoStop}%
\bibitem [{\citenamefont {Kambe}(1969)}]{kambe1969stability}%
  \BibitemOpen
  \bibfield  {author} {\bibinfo {author} {\bibfnamefont {T.}~\bibnamefont
  {Kambe}},\ }\href@noop {} {\bibfield  {journal} {\bibinfo  {journal} {J.
  Phys. Soc. Japan}\ }\textbf {\bibinfo {volume} {26}},\ \bibinfo {pages} {566}
  (\bibinfo {year} {1969})}\BibitemShut {NoStop}%
\bibitem [{\citenamefont {Coenen}\ \emph {et~al.}(2008)\citenamefont {Coenen},
  \citenamefont {Sevilla},\ and\ \citenamefont
  {S{\'a}nchez}}]{coenen2008absolute}%
  \BibitemOpen
  \bibfield  {author} {\bibinfo {author} {\bibfnamefont {W.}~\bibnamefont
  {Coenen}}, \bibinfo {author} {\bibfnamefont {A.}~\bibnamefont {Sevilla}}, \
  and\ \bibinfo {author} {\bibfnamefont {A.}~\bibnamefont {S{\'a}nchez}},\
  }\href@noop {} {\bibfield  {journal} {\bibinfo  {journal} {Phys. Fluids}\
  }\textbf {\bibinfo {volume} {20}},\ \bibinfo {pages} {074104} (\bibinfo
  {year} {2008})}\BibitemShut {NoStop}%
\bibitem [{\citenamefont {Landau}\ and\ \citenamefont
  {Lifshitz}(1959)}]{landau1959fluid}%
  \BibitemOpen
  \bibfield  {author} {\bibinfo {author} {\bibfnamefont {L.~D.}\ \bibnamefont
  {Landau}}\ and\ \bibinfo {author} {\bibfnamefont {E.~M.}\ \bibnamefont
  {Lifshitz}},\ }\href@noop {} {\bibfield  {journal} {\bibinfo  {journal}
  {Course of Theoretical Physics}\ }\textbf {\bibinfo {volume} {4}} (\bibinfo
  {year} {1959})}\BibitemShut {NoStop}%
\bibitem [{\citenamefont {Narasimha}(1985)}]{narasimha1985laminar}%
  \BibitemOpen
  \bibfield  {author} {\bibinfo {author} {\bibfnamefont {R.}~\bibnamefont
  {Narasimha}},\ }\href@noop {} {\bibfield  {journal} {\bibinfo  {journal}
  {Prog. Aerosp. Sci.}\ }\textbf {\bibinfo {volume} {22}},\ \bibinfo {pages}
  {29} (\bibinfo {year} {1985})}\BibitemShut {NoStop}%
\bibitem [{\citenamefont {Pomeau}(2015)}]{pomeau2015transition}%
  \BibitemOpen
  \bibfield  {author} {\bibinfo {author} {\bibfnamefont {Y.}~\bibnamefont
  {Pomeau}},\ }\href@noop {} {\bibfield  {journal} {\bibinfo  {journal}
  {Comptes Rendus M{\'e}canique}\ }\textbf {\bibinfo {volume} {343}},\ \bibinfo
  {pages} {210} (\bibinfo {year} {2015})}\BibitemShut {NoStop}%
\bibitem [{\citenamefont {Pomeau}(1986)}]{pomeau1986front}%
  \BibitemOpen
  \bibfield  {author} {\bibinfo {author} {\bibfnamefont {Y.}~\bibnamefont
  {Pomeau}},\ }\href@noop {} {\bibfield  {journal} {\bibinfo  {journal}
  {Physica D}\ }\textbf {\bibinfo {volume} {23}},\ \bibinfo {pages} {3}
  (\bibinfo {year} {1986})}\BibitemShut {NoStop}%
\bibitem [{\citenamefont {Leweke}\ and\ \citenamefont
  {Provansal}(1995)}]{leweke1995flow}%
  \BibitemOpen
  \bibfield  {author} {\bibinfo {author} {\bibfnamefont {T.}~\bibnamefont
  {Leweke}}\ and\ \bibinfo {author} {\bibfnamefont {M.}~\bibnamefont
  {Provansal}},\ }\href@noop {} {\bibfield  {journal} {\bibinfo  {journal} {J.
  Fluid Mech.}\ }\textbf {\bibinfo {volume} {288}},\ \bibinfo {pages} {265}
  (\bibinfo {year} {1995})}\BibitemShut {NoStop}%
\bibitem [{\citenamefont {Yoda}\ \emph {et~al.}(1994)\citenamefont {Yoda},
  \citenamefont {Hesselink},\ and\ \citenamefont
  {Mungal}}]{yoda1994instantaneous}%
  \BibitemOpen
  \bibfield  {author} {\bibinfo {author} {\bibfnamefont {M.}~\bibnamefont
  {Yoda}}, \bibinfo {author} {\bibfnamefont {L.}~\bibnamefont {Hesselink}}, \
  and\ \bibinfo {author} {\bibfnamefont {M.~G.}\ \bibnamefont {Mungal}},\
  }\href@noop {} {\bibfield  {journal} {\bibinfo  {journal} {J. Fluid Mech.}\
  }\textbf {\bibinfo {volume} {279}},\ \bibinfo {pages} {313} (\bibinfo {year}
  {1994})}\BibitemShut {NoStop}%
\bibitem [{\citenamefont {Danaila}\ \emph {et~al.}(1997)\citenamefont
  {Danaila}, \citenamefont {Du{\v{s}}ek},\ and\ \citenamefont
  {Anselmet}}]{danaila1997coherent}%
  \BibitemOpen
  \bibfield  {author} {\bibinfo {author} {\bibfnamefont {I.}~\bibnamefont
  {Danaila}}, \bibinfo {author} {\bibfnamefont {J.}~\bibnamefont
  {Du{\v{s}}ek}}, \ and\ \bibinfo {author} {\bibfnamefont {F.}~\bibnamefont
  {Anselmet}},\ }\href@noop {} {\bibfield  {journal} {\bibinfo  {journal}
  {Physics of Fluids}\ }\textbf {\bibinfo {volume} {9}},\ \bibinfo {pages}
  {3323} (\bibinfo {year} {1997})}\BibitemShut {NoStop}%
\bibitem [{\citenamefont {Bose}\ and\ \citenamefont
  {Durbin}(2016)}]{bose2016helical}%
  \BibitemOpen
  \bibfield  {author} {\bibinfo {author} {\bibfnamefont {R.}~\bibnamefont
  {Bose}}\ and\ \bibinfo {author} {\bibfnamefont {P.~A.}\ \bibnamefont
  {Durbin}},\ }\href@noop {} {\bibfield  {journal} {\bibinfo  {journal} {Phys.
  Rev. Fluids}\ }\textbf {\bibinfo {volume} {1}},\ \bibinfo {pages} {073602}
  (\bibinfo {year} {2016})}\BibitemShut {NoStop}%
\bibitem [{\citenamefont {Jim{\'e}nez}(1990)}]{jimenez1990transition}%
  \BibitemOpen
  \bibfield  {author} {\bibinfo {author} {\bibfnamefont {J.}~\bibnamefont
  {Jim{\'e}nez}},\ }\href@noop {} {\bibfield  {journal} {\bibinfo  {journal}
  {Journal of Fluid Mechanics}\ }\textbf {\bibinfo {volume} {218}},\ \bibinfo
  {pages} {265} (\bibinfo {year} {1990})}\BibitemShut {NoStop}%
\bibitem [{\citenamefont {Hussein}\ \emph {et~al.}(1994)\citenamefont
  {Hussein}, \citenamefont {Capp},\ and\ \citenamefont
  {George}}]{hussein1994velocity}%
  \BibitemOpen
  \bibfield  {author} {\bibinfo {author} {\bibfnamefont {H.~J.}\ \bibnamefont
  {Hussein}}, \bibinfo {author} {\bibfnamefont {S.~P.}\ \bibnamefont {Capp}}, \
  and\ \bibinfo {author} {\bibfnamefont {W.~K.}\ \bibnamefont {George}},\
  }\href@noop {} {\bibfield  {journal} {\bibinfo  {journal} {J. Fluid Mech.}\
  }\textbf {\bibinfo {volume} {258}},\ \bibinfo {pages} {31} (\bibinfo {year}
  {1994})}\BibitemShut {NoStop}%
\bibitem [{\citenamefont {Das}\ \emph {et~al.}(2017)\citenamefont {Das},
  \citenamefont {Bansal},\ and\ \citenamefont {Manghnani}}]{das2017generation}%
  \BibitemOpen
  \bibfield  {author} {\bibinfo {author} {\bibfnamefont {D.}~\bibnamefont
  {Das}}, \bibinfo {author} {\bibfnamefont {M.}~\bibnamefont {Bansal}}, \ and\
  \bibinfo {author} {\bibfnamefont {A.}~\bibnamefont {Manghnani}},\ }\href@noop
  {} {\bibfield  {journal} {\bibinfo  {journal} {Journal of Fluid Mechanics}\
  }\textbf {\bibinfo {volume} {811}},\ \bibinfo {pages} {138} (\bibinfo {year}
  {2017})}\BibitemShut {NoStop}%
\bibitem [{\citenamefont {Thielicke}\ and\ \citenamefont
  {Stamhuis}(2014)}]{thielicke2014pivlab}%
  \BibitemOpen
  \bibfield  {author} {\bibinfo {author} {\bibfnamefont {W.}~\bibnamefont
  {Thielicke}}\ and\ \bibinfo {author} {\bibfnamefont {E.}~\bibnamefont
  {Stamhuis}},\ }\href@noop {} {\bibfield  {journal} {\bibinfo  {journal}
  {Journal of open research software}\ }\textbf {\bibinfo {volume} {2}}
  (\bibinfo {year} {2014})}\BibitemShut {NoStop}%
\bibitem [{\citenamefont {Keane}\ and\ \citenamefont
  {Adrian}(1990)}]{keane1990optimization}%
  \BibitemOpen
  \bibfield  {author} {\bibinfo {author} {\bibfnamefont {R.~D.}\ \bibnamefont
  {Keane}}\ and\ \bibinfo {author} {\bibfnamefont {R.~J.}\ \bibnamefont
  {Adrian}},\ }\href@noop {} {\bibfield  {journal} {\bibinfo  {journal}
  {Measurement science and technology}\ }\textbf {\bibinfo {volume} {1}},\
  \bibinfo {pages} {1202} (\bibinfo {year} {1990})}\BibitemShut {NoStop}%
\bibitem [{\citenamefont {Raffel}\ \emph {et~al.}(2018)\citenamefont {Raffel},
  \citenamefont {Willert}, \citenamefont {Scarano}, \citenamefont {K{\"a}hler},
  \citenamefont {Wereley},\ and\ \citenamefont
  {Kompenhans}}]{raffel2018particle}%
  \BibitemOpen
  \bibfield  {author} {\bibinfo {author} {\bibfnamefont {M.}~\bibnamefont
  {Raffel}}, \bibinfo {author} {\bibfnamefont {C.~E.}\ \bibnamefont {Willert}},
  \bibinfo {author} {\bibfnamefont {F.}~\bibnamefont {Scarano}}, \bibinfo
  {author} {\bibfnamefont {C.~J.}\ \bibnamefont {K{\"a}hler}}, \bibinfo
  {author} {\bibfnamefont {S.~T.}\ \bibnamefont {Wereley}}, \ and\ \bibinfo
  {author} {\bibfnamefont {J.}~\bibnamefont {Kompenhans}},\ }\href@noop {}
  {\emph {\bibinfo {title} {Particle image velocimetry: a practical guide}}}\
  (\bibinfo  {publisher} {Springer},\ \bibinfo {year} {2018})\BibitemShut
  {NoStop}%
\bibitem [{Note1()}]{Note1}%
  \BibitemOpen
  \bibinfo {note} {Global conservation of momentum for the base flow implies
  that the Reynolds number based on $U_{cl}, y_{1/4}$ is identical to the
  Reynolds number based on $U_{av}, D$.}\BibitemShut {Stop}%
\bibitem [{\citenamefont {Trefethen}(2000)}]{trefethen2000spectral}%
  \BibitemOpen
  \bibfield  {author} {\bibinfo {author} {\bibfnamefont {L.~N.}\ \bibnamefont
  {Trefethen}},\ }\href@noop {} {\emph {\bibinfo {title} {Spectral Methods in
  MATLAB}}}\ (\bibinfo  {publisher} {SIAM},\ \bibinfo {year}
  {2000})\BibitemShut {NoStop}%
\bibitem [{\citenamefont {Xie}\ and\ \citenamefont
  {Lin}(2009)}]{xie2009efficient}%
  \BibitemOpen
  \bibfield  {author} {\bibinfo {author} {\bibfnamefont {M.}~\bibnamefont
  {Xie}}\ and\ \bibinfo {author} {\bibfnamefont {J.}~\bibnamefont {Lin}},\
  }\href@noop {} {\bibfield  {journal} {\bibinfo  {journal} {International
  journal for numerical methods in fluids}\ }\textbf {\bibinfo {volume} {61}},\
  \bibinfo {pages} {780} (\bibinfo {year} {2009})}\BibitemShut {NoStop}%
\bibitem [{\citenamefont {Kulkarni}\ and\ \citenamefont
  {Agarwal}(2007)}]{kulkarni2007viscous}%
  \BibitemOpen
  \bibfield  {author} {\bibinfo {author} {\bibfnamefont {T.~M.}\ \bibnamefont
  {Kulkarni}}\ and\ \bibinfo {author} {\bibfnamefont {A.}~\bibnamefont
  {Agarwal}},\ }\href@noop {} {\bibfield  {journal} {\bibinfo  {journal} {ISVR
  Report No. 317}\ } (\bibinfo {year} {2007})}\BibitemShut {NoStop}%
\end{thebibliography}%

\end{document}